\documentclass[floatfix,showpacs]{revtex4}
\usepackage{eurosym}
\usepackage[utf8]{inputenc}
\usepackage{amsmath}
\usepackage{amsfonts}
\usepackage{t4phonet}
\usepackage{amssymb}
\usepackage{setspace}
\usepackage{tipa}
\usepackage{epsfig}
\usepackage{epstopdf}
\usepackage{graphicx}
\usepackage{caption}
\usepackage{float}
\usepackage{subfigure}

\begin{document}

\title{$\mathcal{CP}$ symmetry in optical systems }
\author{Brenda Dana, Alon Bahabad and Boris A. Malomed}
\address{Department of Physical Electronics, School of Electrical
Engineering, Fleischman Faculty of Engineering, Tel-Aviv University,
Tel-Aviv 69978, Israel}

\begin{abstract}
We introduce a model of a dual-core optical waveguide with opposite signs of
the group-velocity-dispersion (GVD) in the two cores, and a phase-velocity
mismatch between them. The coupler is embedded into an active host medium,
which provides for the linear coupling of a gain-loss type between the two
cores. The same system can be derived, without phenomenological assumptions,
by considering the three-wave propagation in a medium with the quadratic
nonlinearity, provided that the depletion of the second-harmonic pump is
negligible. This linear system offers an optical realization of the
charge-parity ($\mathcal{CP}$) symmetry, while the addition of the
intra-core cubic nonlinearity breaks the symmetry. By means of direct
simulations and analytical approximations, it is demonstrated that the
linear system generates expanding Gaussian states, while the nonlinear one
gives rise to broad oscillating solitons, as well as a general family of
stable stationary gap solitons.
\end{abstract}

\pacs{05.45.Yv; 42.65.Tg; 11.30.Er; 42.79.Gn}
\maketitle

\section{Introduction}

Charge-parity-time ($\mathcal{CPT}$) symmetry is the most fundamental type
of symmetry in quantum field theory \cite%
{1742-6596-335-1-012011,AguilarArevalo20131303}, where it holds for all
relativistically invariant systems obeying the causality principle. Its
reduced form, \textit{viz}., the $\mathcal{CP}$ symmetry, is almost exact
too, save the small violation by weak nuclear forces \cite{particle}. The $%
\mathcal{CPT}$ operator is composed of three factors : parity
transformation, $\mathcal{P}$, which reverses the coordinate axes; charge
conjugation, $\mathcal{C}$, which swaps particles and antiparticles; and
time reversal, $\mathcal{T}$.

The proof of the $\mathcal{CPT}$ and $\mathcal{CP}$ symmetries (when the
latter is relevant) applies to Hermitian Hamiltonians ($H$), subject to the
condition $H=H^{\dagger }$, which guarantees that the spectrum of the
Hamiltonian is real. However, one cannot deduce from the $\mathcal{CPT}$ or $%
\mathcal{CP}$ symmetry that the respective Hamiltonian is necessarily
Hermitian \cite{Ham}. Indeed, the consideration of Hamiltonians which
commute with a reduced symmetry operator, $\mathcal{PT}$ , demonstrates that
they may contain an anti-Hermitian (dissipative) part, provided that it is
spatially antisymmetric (odd), while the Hermitian one is even \cite{Bender}%
. The spectrum of such a Hamiltonian remains purely real up to a critical
value of the strength of the anti-Hermitian part, at which the $\mathcal{PT}$
symmetry is broken, making the system an essentially dissipative one
(recently, a model with \textit{unbreakable} $\mathcal{PT}$ symmetry was
found; it includes defocusing cubic nonlinearity with the local strength
growing from the center to periphery \cite{unbreakable}).

While in the quantum theory the possibility of the existence of
non-Hermitian $\mathcal{PT}$-symmetric Hamiltonians is a purely theoretical
one, such systems have been realized, theoretically \cite%
{Muga,Muga1,Muga2,Muga3,Muga4,Muga5,review} and experimentally \cite%
{Kip,Feng05082011,citeulike:11031904,Ruter:09,10.1117/12.807739,PhysRevA.82.010103,2012Natur.488..163R}%
, in optics, making use of the fact that the wave-propagation equation,
derived in the standard paraxial approximation, is identical to the Schr\"{o}%
dinger equation in nonrelativistic quantum mechanics. In this context, the
spatially symmetric and antisymmetric Hermitian and anti-Hermitian terms of
the Hamiltonian are represented, respectively, by even and odd distributions
of the refractive index, and of the local gain-loss coefficient in the
photonic medium. A $\mathcal{PT}$-symmetric electronic circuit was built
too, following similar principles \cite{Kottos}.

The essential role played by the Kerr nonlinearity in optics has suggested
the development of models in which the Hamiltonian includes a quartic
Hermitian part too. The nonlinearity gives rise to families of $\mathcal{PT}$
-symmetric solitons, that were investigated in detail in continuous and
discrete systems \cite%
{Muga5,PTsolitons,PTsolitons1,PTsolitons2,PTsolitons3,PTsolitons4,PTsolitons5}%
, including $\mathcal{PT}$ -symmetric dual-core couplers \cite%
{couplers,couplers1,coupler-management}. Models combining the $\mathcal{PT}$
symmetry with quadratic nonlinearity in the dynamical equations (i.e., cubic
terms in the respective Hamiltonians) were also elaborated \cite%
{VVK1,VVK2,VVK3}.

As a subject of quantum field theory, the $\mathcal{CPT}$ and $\mathcal{CP}$
symmetries mainly relate to elementary particles \cite%
{2012arXiv1201.1594S,srednicki2007quantum,kursunogammalu2013confluence,2013PhLB..718.1500C}%
. On the other hand, the above-mentioned works on the implementation of
non-Hermitian $\mathcal{PT}$-symmetric Hamiltonians in photonics suggest
looking for a possibility to design optical settings that would realize
non-Hermitian Hamiltonians featuring the full $\mathcal{CPT}$ symmetry, as
well as its $\mathcal{CP}$ reduction. A possibility to implement the former
symmetry was recently explored in Ref. \cite{something}, which addressed not
optics, but rather a two-component Bose-Einstein condensate with the
spin-orbit coupling between the components, one of which is subject to the
action of loss, and the other one is supported by gain. In terms of optics
systems, the symmetry of that models is similar to the $\mathcal{PT}$
symmetry in a dual-core waveguide, with a combination of continuous $%
\mathcal{P}$\ transformation acting in the longitudinal direction, and
another $\mathcal{P}$\ transformation which swaps the two cores. A similar
symmetry was proposed in Ref. \cite{coupler-management}, which put forward a
$\mathcal{PT}$-symmetric coupler subject to the action of \textquotedblleft
management", in the form of periodic simultaneous switch of the signs of the
coupling and gain-loss coefficients.

The present work aims to offer emulation of the $\mathcal{CP}$ symmetry in a
two-component optical system, which, at the phenomenological level, may be
considered as a dual-core waveguide with opposite signs of the
group-velocity dispersion (GVD) in the cores and a phase-velocity mismatch
between them, embedded into an active medium. We demonstrate that the system
can be derived, without phenomenological assumptions, as a model of the
spatial-domain propagation for two fundamental-frequency (FF) components
with orthogonal polarizations of light, pumped by an undepleted
second-harmonic (SH) wave in a birefringent medium with the $\chi ^{(2)}$
nonlinearity. We\ further investigate conditions for persistence and
breaking of the $\mathcal{CP}$ symmetry, both analytically and numerically.
In particular, the addition of cubic (Kerr, alias $\chi ^{(3)}$) terms to
the system with the active coupling breaks the symmetry at the nonlinear
level, but helps to stabilize confined breather states (oscillatory
solitons), and gives rise to a family of stable stationary gap solitons.

The paper is organized as follows: Section 2 introduces the model, in its
both forms (phenomenological and the one based on the $\chi ^{(2)}$
interaction) and reports analytical results. Numerical findings for the
linear and nonlinear systems are presented in Section 3, and Section 4
concludes the paper.

\section{The model and analytical results}

\subsection{The system: phenomenological formulation}

At the phenomenological level, we consider the copropagation of optical
modes $u$ and $v$ in a dual-core coupler with opposite GVD signs in the
cores (cf. Refs. \cite{Kaup,Kaup1,Dana:14}, where a similar feature was
introduced in different contexts, and also Refs. \cite{Maim1,Maim2,Maim3},
where systems with opposite signs of group velocities were considered in the
contexts of coupled right- and left-handed waveguides), and a phase-velocity
mismatch $2q$ between them. The linear coupling between the cores is
provided by cross-gain terms, with strength $\gamma $, which is possible
when the coupler is embedded into an active medium, as recently proposed in
Ref. \cite{Barash}. The model is represented by the following system of
propagation equations, which include the Kerr nonlinearity too, with
respective coefficient $\sigma $ (all the quantities are dimensionless):
\begin{eqnarray}
iu_{z}+(1/2)u_{tt}-qu+\sigma |u|^{2}u &=&i\gamma v,  \label{u} \\
iv_{z}-(1/2)v_{tt}+qv+\sigma |v|^{2}v &=&i\gamma u.  \label{v}
\end{eqnarray}%
Here $z$ is the propagation distance, $t$ is the reduced time \cite%
{agrawal2001nonlinear}, the GVD coefficients are scaled to be $\pm 1$, and $%
\gamma <0$ may be transformed to $\gamma >0$ by changing $v\rightarrow -v$.
Positive and negative values of $\sigma $ can also be transformed into each
other by substitution $\left( u^{\ast },v^{\ast }\right) \equiv \left(
\tilde{v},\tilde{u}\right) $, therefore, in what follows below we consider
only $\sigma >0$. Then, rescaling allows one to fix $\sigma \equiv 1$, but
we prefer to keep it as a free parameter, the variation of which helps to
monitor a transition from the weakly nonlinear system to a strongly
nonlinear one.

It is relevant to mention that a dissipative discrete system with opposite
signs of the discrete dispersion and a wavenumber mismatch between the
components was introduced in Ref. \cite{VVK}. However, that model included
the dissipative coefficient in a single equation, therefore it did not
realize the symmetry considered here.

Equations (\ref{u}) and (\ref{v}) can be derived from the non-Hermitian
(complex) Lagrangian, which is usual for $\mathcal{PT}$-symmetric systems:%
\begin{gather}
L=\int_{-\infty }^{+\infty }\left[ i\left( u^{\ast }u_{z}+v^{\ast
}v_{z}\right) +\frac{1}{2}\left( \left\vert v_{t}\right\vert -\left\vert
u_{t}\right\vert ^{2}\right) +q\left( |v|^{2}-|u|^{2}\right) +\frac{\sigma }{%
2}\left( |u|^{4}+|v|^{4}\right) \right] dt  \notag \\
-i\gamma \int_{-\infty }^{+\infty }\left( u^{\ast }v+uv^{\ast }\right) dt,
\label{L}
\end{gather}%
which generates the respective non-Hermitian Hamiltonian in an obvious way.\
The total energy,
\begin{equation}
E(z)=\int_{-\infty }^{+\infty }\left[ |u\left( z,t)\right) |^{2}+|v\left(
z,t\right) |^{2}\right] dt\equiv E_{u}(z)+E_{v}(z),  \label{P}
\end{equation}%
is not conserved by Eqs. (\ref{u}) and (\ref{v}). Instead, the system gives
rise to the following energy-balance equations:%
\begin{equation}
\frac{dE_{u}}{dz}=\frac{dE_{v}}{dz}\equiv \frac{1}{2}\frac{dE}{dz}=\gamma
\int_{-\infty }^{+\infty }\left( uv^{\ast }+u^{\ast }v\right) dt.
\label{dP/dz2}
\end{equation}%
The equality of $dE_{u}/dz$ and $dE_{v}/dz$, i.e., the conservation of $%
E_{u}-E_{v}$, means that the linear coupling of the present type causes
mutual amplification or attenuation of both components.

The linear version of Eqs. (\ref{u}) and (\ref{v}), with $\sigma =0$, are
invariant with respect to the $\mathcal{CP}$ transformation, defined as%
\begin{equation}
\left( u,v\right) \rightarrow \left( \tilde{u}\equiv v^{\ast },\tilde{v}%
\equiv u^{\ast }\right) ,  \label{CPT}
\end{equation}%
where the swap of $u$ and $v$ stands for $\mathcal{P}$, and the complex
conjugation -- for $\mathcal{C}$ (conserved $E_{u}-E_{v}$ may be considered
as the respective charge). It is relevant to compare the system of Eqs. (\ref%
{u}), (\ref{v}) and their invariance transformation (\ref{CPT}) with the
previously studied model of the $\mathcal{PT}$-symmetric coupler, which was
based on the following equations \cite%
{couplers,couplers1,coupler-management}:
\begin{gather}
iu_{z}+(1/2)u_{tt}+\sigma |u|^{2}u=i\gamma v,  \label{uPT} \\
iv_{z}+(1/2)v_{tt}+\sigma |v|^{2}v=-i\gamma v.  \label{vPT}
\end{gather}%
Obviously, Eqs. (\ref{uPT}) and (\ref{vPT}) are invariant with respect to
transformation $\left( u,v,z\right) \rightarrow \left( \tilde{u}\equiv
v^{\ast },\tilde{v}\equiv u^{\ast },\tilde{z}\equiv -z\right) $, which, in
the present context, may be considered as corresponding to the $\mathcal{CPT}
$ symmetry, the reversal of $z$ playing the role of additional $\mathcal{T}$.

It is relevant too to compare the present model to the system of equations
with opposite GVD terms, coupled by the usual conservative terms, rather
than by those representing the gain and loss \cite{Dana:14}:
\begin{equation}
\begin{array}{c}
iu_{z}+(1/2)u_{tt}+\sigma |u|^{2}u+Kv=0, \\
iv_{z}-(1/2)v_{tt}+\sigma |v|^{2}v+Ku=0,%
\end{array}
\label{K}
\end{equation}%
where $K$ is a real coupling constant. The linear version of this system is
invariant with respect to the \emph{anti}-$\mathcal{CP}$ transformation: $%
\left( u,v\right) \rightarrow \left( \tilde{u}\equiv v^{\ast },\tilde{v}%
\equiv -u^{\ast }\right) ,$ \textquotedblleft anti" corresponding to the
relative sign flip, cf. Eq. (\ref{CPT}).

The nonlinearity breaks the symmetry of system (\ref{u}), (\ref{v}), as the
opposite relative signs of the GVD and cubic terms in the two equations make
it impossible to swap $u$ and $v$, which represents the $\mathcal{P}$
transformation in Eq. (\ref{CPT}). Nevertheless, nonlinear effects are
obviously interesting too. It is demonstrated below that the nonlinearity
creates solitons in the present system. In this connection, it is relevant
to mention recently introduced nonlinear models with alternating gain and
loss, which do not obey the condition of the $\mathcal{PT}$ symmetry, but
nevertheless support stable solitons \cite{non-PT1,non-PT2,non-PT3}.

A solution to the linear version of Eqs. (\ref{u}-\ref{v}) in the form of
plane waves, $\left\{ u,v\right\} =\left\{ u_{0},v_{0}\right\} \exp \left(
ikz-i\omega t\right) ,$ produces a dispersion relation for the wavenumber
and frequency:%
\begin{equation}
k=\pm \sqrt{\left( q+\frac{1}{2}\omega ^{2}\right) ^{2}-\gamma ^{2}}.
\label{DR2}
\end{equation}%
Obviously, in the case of $q>0$ the spectrum given by Eq. (\ref{DR2}) is
pure real, provided that%
\begin{equation}
\left\vert \gamma \right\vert <\gamma _{\mathrm{thr}}\equiv q,  \label{thr}
\end{equation}%
while in the case of $q<0$ the spectrum always includes an imaginary
component. The change of the spectrum from real to a partly imaginary one,
with the increase of the gain-loss coefficient, at $|\gamma |=q$ (provided
that $q>0$) implies the breakup of the $\mathcal{CP}$ symmetry, similar to
the phase transition which is the generic feature of $\mathcal{PT}$%
-symmetric systems \cite{nature}. If condition (\ref{thr}) holds, the
spectrum given by Eq. (\ref{DR2}) features a \textit{bandgap},%
\begin{equation}
k^{2}<q^{2}-\gamma ^{2}.  \label{gap}
\end{equation}

\subsection{The linear model: physical derivation}

While the system of Eqs. (\ref{u}), (\ref{v}) was introduced above
phenomenologically, its linear version can be derived, in the spatial domain
(rather than in the temporal one), starting from the fundamental propagation
model for two FF and one SH\ components of light waves, $u$, $\hat{v}$ and $w
$, respectively, in the dissipation-free medium with the Type-II $\chi ^{(2)}
$ interaction \cite{chi2-0,chi2-1,chi2-2,chi2-3}:%
\begin{gather}
iu_{z}+(1/2)u_{xx}-qu=-\hat{v}^{\ast }w,  \label{chi2u} \\
i\hat{v}_{z}+(1/2)\hat{v}_{xx}-q\hat{v}=-u^{\ast }w,  \label{chi2v} \\
2iw_{z}+(1/2)w_{xx}=-\left( 1/2\right) u\hat{v},  \label{chi2w}
\end{gather}%
where $x$ is the transverse coordinate, and $q$ is an FF-SH wavenumber
mismatch. Then, adopting the usual approximation for parametric down-conversion, of undepleted SH pump,
we replace it by a constant, $w=-i\gamma $, neglecting Eq. (\ref{chi2w}),
denote $\hat{v}^{\ast }\equiv v$, and apply the complex conjugation to Eq. (%
\ref{chi2v}):%
\begin{gather}
iu_{z}+(1/2)u_{xx}-qu=i\gamma v,  \label{uphys} \\
iv_{z}-(1/2)v_{xx}+qv=i\gamma u.  \label{vphys}
\end{gather}%
These equations differ from the linear version of Eqs. (\ref{u}), (\ref{v})
only by the replacement of $t$ by $x$.

As for cubic terms, they can be added to Eqs. (\ref{chi2u}-\ref{chi2w}) as
ones accounting for the Kerr nonlinearity in the $\chi ^{(2)}$ waveguide.
However, in terms of Eqs. (\ref{uphys}) and (\ref{vphys}), the resulting
cubic terms will be different from those adopted in Eqs. (\ref{u}) and (\ref%
{v}), as the above-mentioned complex conjugation of Eq. (\ref{chi2v}) will
produce the cubic term in Eq. (\ref{vphys}) with the sign opposite to that
in Eq. (\ref{v}) [incidentally, in this case the cubic terms do not break
the $\mathcal{CP}$ invariance of Eqs. (\ref{u}) and (\ref{v})]. Furthermore,
because the original components, $u$ and $\hat{v}$, correspond to two
orthogonal polarizations of the FF wave, the nonlinear extension of Eqs. (%
\ref{uphys}) and (\ref{vphys}) should also include the respective XPM
(cross-phase-modulation) terms, \textit{viz}., $(2/3)\sigma |v|^{2}u$ and $%
-(2/3)\sigma |u|^{2}v$, respectively, assuming that the four-wave mixing
terms may be neglected, as usual, due to sufficiently strong birefringence
\cite{agrawal2001nonlinear}. In the present work, we focus on the nonlinear
terms adopted in Eqs. (\ref{u}), (\ref{v}), while those corresponding to the
derivation for the $\chi ^{(2)}$ system will be considered elsewhere.

\subsection{The analytical approximation for broad pulses}

The system based on Eqs. (\ref{u}), (\ref{v}) can be investigated in an
analytical form for broad small-amplitude pulses, with widths ($\tau $) and
amplitudes satisfying conditions
\begin{equation}
\tau ^{2}\gg 1/q;~U_{0}^{2},V_{0}^{2}\ll 1/\left( \sigma q\right) .
\label{broad}
\end{equation}%
In this case, the linearized version of the system yields, in the lowest
approximation, two different relations between the field components: one
solution has%
\begin{equation}
v\left( z,t\right) =\frac{i\gamma }{q+\sqrt{q^{2}-\gamma ^{2}}}%
u(z,t),~u(z,t)=e^{-i\sqrt{q^{2}-\gamma ^{2}}z}~\tilde{u}(z,t),  \label{v-u}
\end{equation}%
and another one features%
\begin{equation}
u\left( z,t\right) =\frac{-i\gamma }{q+\sqrt{q^{2}-\gamma ^{2}}}%
v(z,t),~v(z,t)=e^{i\sqrt{q^{2}-\gamma ^{2}}z}~\tilde{v}(z,t),  \label{u-v}
\end{equation}%
where $\tilde{u}(z,t)$ and $\tilde{v}(z,t)$ are slowly varying amplitudes,
in comparison with $\exp \left( \pm i\sqrt{q^{2}-\gamma ^{2}}z\right) $. The
substitution of these expressions into Lagrangian (\ref{L}) leads to a real
effective Lagrangians for the slowly varying functions (its imaginary part
cancels out in the present approximation), which, in turn, give rise to
either one of the two following nonlinear Schr\"{o}dinger (NLS) equations
for the slow evolution:%
\begin{equation}
i\frac{\partial }{\partial z}\left(
\begin{array}{c}
\tilde{u} \\
\tilde{v}%
\end{array}%
\right) \pm \frac{1}{2}D_{\mathrm{eff}}\frac{\partial ^{2}}{\partial t^{2}}%
\left(
\begin{array}{c}
\tilde{u} \\
\tilde{v}%
\end{array}%
\right) +\sigma _{\mathrm{eff}}\left(
\begin{array}{c}
\left\vert \tilde{u}\right\vert ^{2}\tilde{u} \\
\left\vert \tilde{v}\right\vert ^{2}\tilde{v}%
\end{array}%
\right) =0,  \label{slow}
\end{equation}%
where $+$ and $-$ pertain to $\tilde{u}$ and $\tilde{v}$, respectively,
while the effective GVD and nonlinearity coefficients are
\begin{equation}
D_{\mathrm{eff}}=\frac{\sqrt{q^{2}-\gamma ^{2}}}{q},~\sigma _{\mathrm{eff}%
}=\sigma \frac{2q^{2}-\gamma ^{2}}{q\left( q+\sqrt{q^{2}-\gamma ^{2}}\right)
}~.  \label{eff}
\end{equation}%
Note that Eq. (\ref{slow}) implies that the system conserves the total
energy in the present approximation, which complies with the fact that the
substitution of relations (\ref{v-u}) and (\ref{u-v}) into energy-balance
equations (\ref{dP/dz2}) yields $dE_{u,v}/dz=0$.

Fundamental solutions to the linear version of Eq. (\ref{slow}) are well
known in the form of expanding Gaussians (coherent states, in terms of
quantum mechanics) \cite{agrawal2001nonlinear}:%
\begin{equation}
\left(
\begin{array}{c}
\tilde{u} \\
\tilde{v}%
\end{array}%
\right) =\left(
\begin{array}{c}
U_{0} \\
V_{0}%
\end{array}%
\right) \frac{1}{\sqrt{t_{0}^{2}\pm iD_{\mathrm{eff}}z}}\exp \left( -\frac{%
t^{2}}{2\left( t_{0}^{2}\pm iD_{\mathrm{eff}}z\right) }\right) ,
\label{Gauss}
\end{equation}%
where $t_{0}$ is an initial width, and $\left( U_{0},V_{0}\right) $ are
arbitrary amplitudes. This result is drastically different from that
obtained for broad pulses in the linear version of Eq. (\ref{K}) with the
conservative coupling \cite{Dana:14}. The latter system gives rise to an
effective equation for a slowly varying function in the form of a single
linear Schr\"{o}dinger equation with the GVD term periodically (in $z$)
changing its sign, thus generating robust oscillating Gaussian pulses,
rather than expanding ones (\ref{Gauss}).

Further, the full nonlinear equations (\ref{slow}) for $\tilde{u}$ and $%
\tilde{v}$ give rise to commonly known solutions for bright and dark
solitons, respectively. In particular, the bright NLS solitons with an
arbitrary (small) propagation constant, $0<\kappa \ll q$, are
\begin{equation}
\tilde{u}_{\mathrm{sol}}=\sqrt{2\kappa /\sigma _{\mathrm{eff}}}e^{i\kappa z}%
\mathrm{sech}\left( \sqrt{2\kappa /D_{\mathrm{eff}}}t\right) .  \label{sol}
\end{equation}

It is also worthy to note that the equations for $\tilde{u}$ and $\tilde{v}$
are, severally, Galilean invariant ones, i.e., the linear and nonlinear
versions of the equation for $\tilde{u}$ give rise, severally, to moving
Gaussians and bright solitons, while the full underlying system (\ref{u}), (%
\ref{v}) does not feature the Galilean invariance.

\subsection{Gap solitons}

Inside of bandgap (\ref{gap}), it is natural to look for stationary
gap-soliton solutions \cite{Sterke} of Eqs. (\ref{u}), (\ref{v}) in the form
of
\begin{equation}
\left\{ u,v\right\} =e^{ikz}\left\{ U(t),V(t)\right\} ,  \label{stat}
\end{equation}%
for which Eqs. (\ref{u}) and (\ref{v}) reduce to a system of stationary
equations,%
\begin{eqnarray}
-kU+\frac{1}{2}\frac{d^{2}U}{dt^{2}}-qU+|U|^{2}U &=&i\gamma V,  \label{U} \\
-kV-\frac{1}{2}\frac{d^{2}V}{dt^{2}}+qV+|V|^{2}V &=&i\gamma U  \label{V}
\end{eqnarray}%
(recall $\sigma =+1$ is fixed). Unlike the usual coupled-mode system for
Bragg gratings \cite{Sterke}, there is no substitution which could reduce
Eqs. (\ref{U}) and (\ref{V}) to a single equation, therefore the system
should be solved numerically, in the general case. The stability of the gap
solitons should then be tested numerically too. As concerns broad solitons (%
\ref{sol}), they actually correspond to the gap solitons at values of $k$
close to the bottom of bandgap (\ref{gap}), $k=-\sqrt{q^{2}-\gamma ^{2}}%
+\kappa $, cf. a similar relation between the general gap solitons and broad
ones in the standard model of nonlinear Bragg gratings \cite{broad-Bragg}.

\begin{figure}[tbp]
\centering
\includegraphics[width=0.5\textwidth]{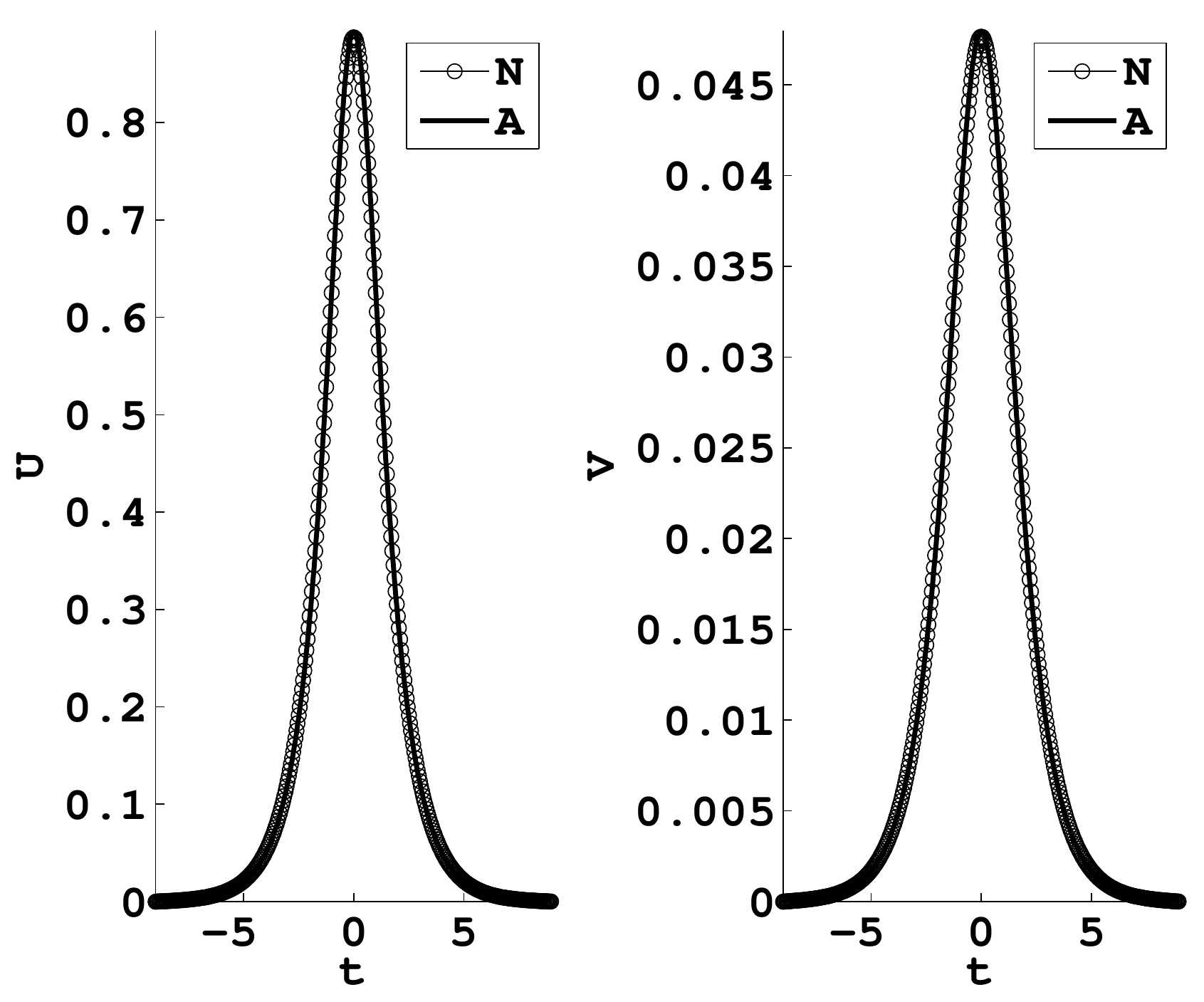}
\caption{Numerical and analytical solutions for the two components of a
strongly asymmetric gap soliton, $U(t)$ and $V(t)$, obtained from Eqs. (%
\protect\ref{U}) and (\protect\ref{V}) for $q=1$, $\protect\sigma =1$, $%
\protect\gamma =0.1$, and $k=-(3/5)q$. The respective analytical solution is
given by Eqs. (\protect\ref{0}) and (\protect\ref{3/5}). }
\label{fig:c_gamma_0.1_k_3_5}
\end{figure}

A gap-soliton solution to Eqs. (\ref{U}) and (\ref{V}) with strong asymmetry
between the two components can be found in an approximate form for the limit
case of weak gain-loss coupling, $\gamma ^{2}\ll q^{2}$, at $k=0$, i.e.,
exactly at the central point of bandgap (\ref{gap}). The zero-order
approximation (for $\gamma =0$) is%
\begin{equation}
U=\sqrt{2\left( q+k\right) }\mathrm{sech}\left( \sqrt{2\left( q+k\right) }%
t\right) ,~V=0,  \label{0}
\end{equation}%
for $q+k>0$ (in this approximation, $k=0$ is not required). Then, the
first-order correction is determined by the linearized equation for $V$:%
\begin{equation}
\left( q-k\right) V-\frac{1}{2}\frac{d^{2}V}{dt^{2}}=i\gamma \sqrt{2\left(
q+k\right) }\mathrm{sech}\left( \sqrt{2\left( q+k\right) }t\right) .
\label{1}
\end{equation}%
An exact closed-form solution to Eq. (\ref{1}) can be found, by means of the
Fourier transform, at the center of the bandgap, i.e., at $k=0$ (a similar
solution was reported, in another context, in Ref. \cite{Roy}):%
\begin{equation}
V(t)=i\gamma \sqrt{2q}\left\{ \sqrt{2q}t\exp \left( -\sqrt{2q}t\right)
+\cosh \left( \sqrt{2q}t\right) \ln \left[ 1+\exp \left( -2\sqrt{2q}t\right) %
\right] \right\} .  \label{Vexact}
\end{equation}%
An exact solution to Eq. (\ref{1}) is available too in the special case of $%
k=-(3/5)q<0$:%
\begin{gather}
V(t)=i\gamma \frac{\sqrt{5}\exp \left( {-{4\sqrt{\frac{q}{5}}t}}\right) }{2%
\sqrt{q}}\left\{ {\exp }\left( {2}\sqrt{\frac{q}{5}}t\right) {+\exp }\left( {%
6}\sqrt{\frac{q}{5}}t\right) \right.   \notag \\
\left. {-\exp }\left( {8}\sqrt{\frac{q}{5}}t\right) \arctan {\left[ \exp
\left( {-2}\sqrt{\frac{q}{5}}t\right) \right] -}\arctan {\left[ \exp \left( {%
2}\sqrt{\frac{q}{5}}t\right) \right] }\right\} .  \label{3/5}
\end{gather}%
These solutions exist under exactly the same condition, $q>0$, which was
adopted above. Although it may not be immediately obvious, both solutions (%
\ref{Vexact}) and (\ref{3/5}) are even functions of $t$, exponentially
decaying at $|t|\rightarrow \infty $. These approximate analytical solutions
are compared with their numerically found counterparts below, see Figs. \ref%
{fig:c_gamma_0.1} and \ref{fig:c_gamma_0.1_k_3_5}.

\section{Numerical results}

\subsection{The linear system}

Equations (\ref{u}-\ref{v}) were solved numerically by means of the
split-step Fourier-transform method \cite{agrawal2001nonlinear}, for four
different sets of initial conditions. Two of them were taken with a Gaussian
pulse in either component:
\begin{eqnarray}
u(z &=&0,t)=\exp ({-0.05t^{2}/2}),~v(z=0,t)=0,  \label{IC1} \\
u(z &=&0,t)=0,~v(z=0,t)=\exp ({-0.05t^{2}/2}),  \label{IC2}
\end{eqnarray}%
\noindent and two other initial sets are given by Eqs. (\ref{v-u}-\ref{u-v})
and (\ref{Gauss}) with $z=0$. The interval for the temporal variable was
fixed as $-800\leq t\leq 800$, to ensure that reflections from its
boundaries did not affect the results, moreover all quantities plotted are dimensionless.

We start the analysis with the case of weak coupling, $0<\gamma \ll q\equiv 1
$. A typical numerical solution for this case, presented in Fig. \ref%
{fig:l_amp_gamma_0.1}, shows the expansion of the Gaussian launched in the
form of initial conditions (\ref{IC1}). The $v$ component remains weak as
the coupling constant is small and, accordingly, the total energy remains
very close to the initial value. A detailed comparison with Eqs. (\ref{v-u}%
-\ref{slow}) demonstrates that the asymptotic stage of the evolution, for
broad pulses, is accurately predicted by the analytical approximation.

\begin{figure}[tbp]
\centering
\subfigure[]{
\includegraphics[width=0.5\textwidth]{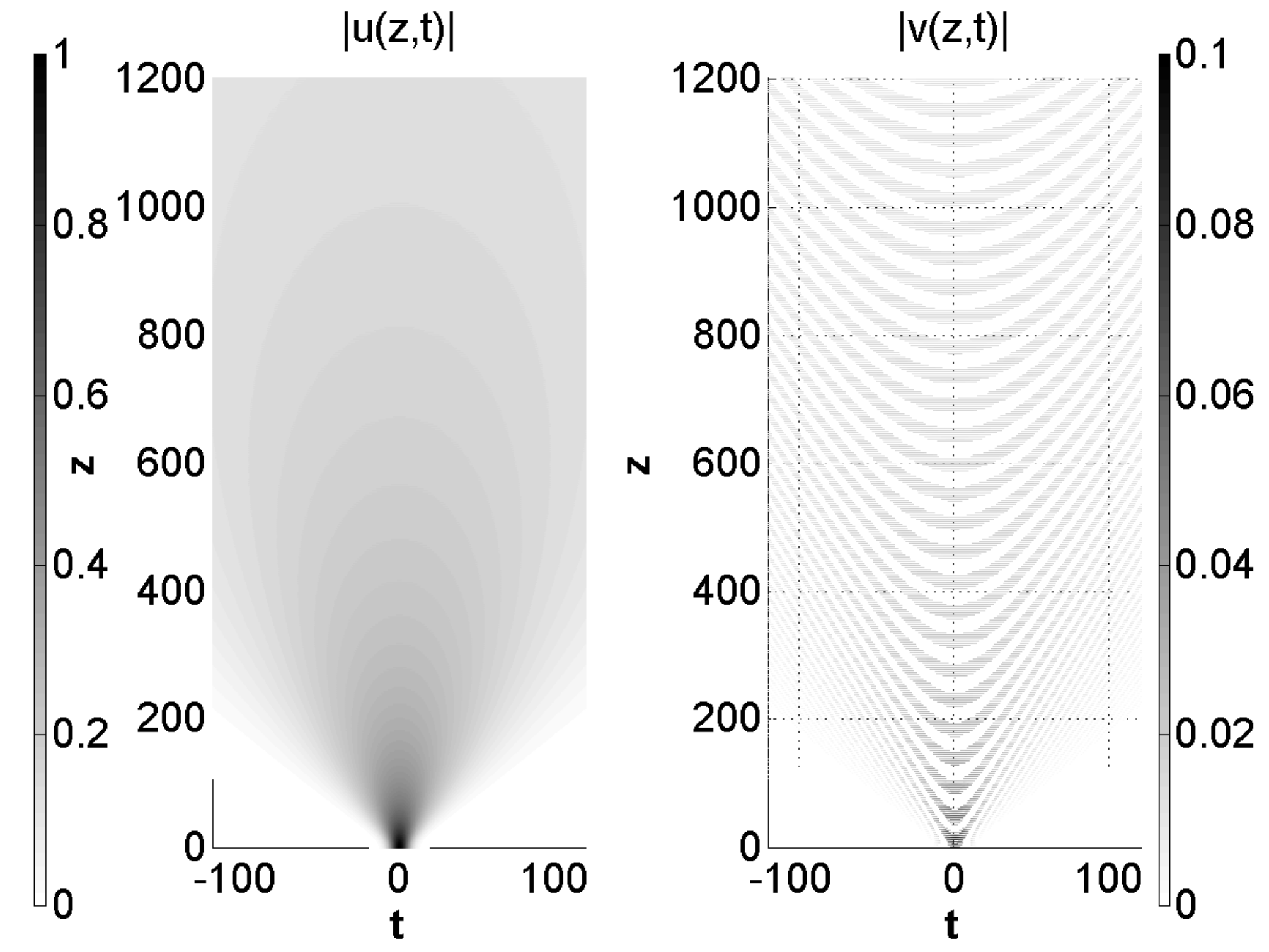}}
\subfigure[]{
\includegraphics[width=0.5\textwidth]{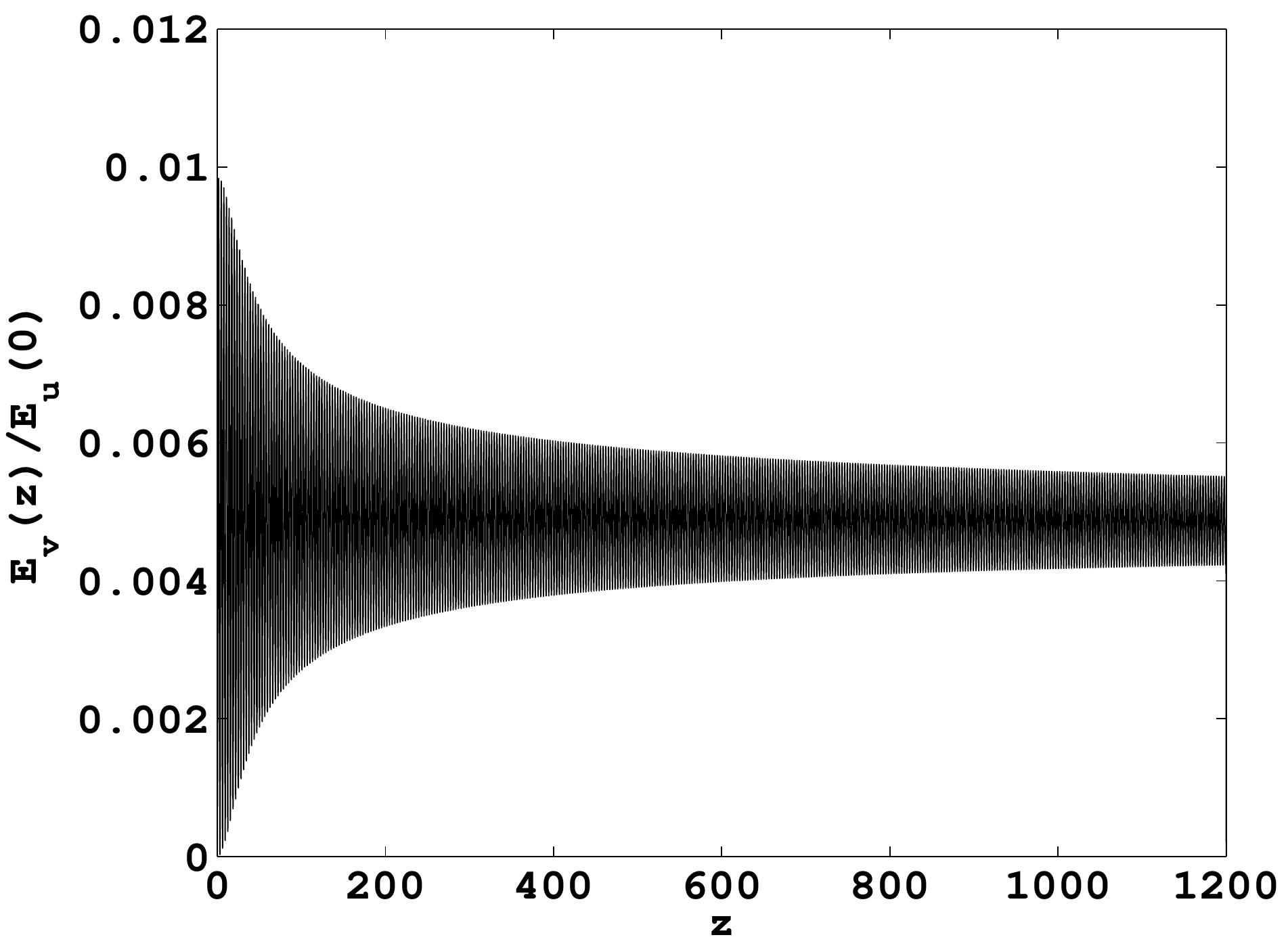}}
\caption{(a) Absolute values $|u(z,t)|$ and $|v(z,t)|$, as functions of the
propagation distance, $z$, and temporal coordinate, $t$, obtained from the
numerical solution of the linear version of Eqs. (\protect\ref{u}) and (%
\protect\ref{v}) with initial conditions (\protect\ref{IC1}), in the case of
weak coupling, $\protect\gamma =0.1$, $q=1$. (b) The evolution of the
integral energy of the $v$ component, defined as per Eq. (\protect\ref{P})
and normalized to the initial energy. The evolution of energy of the $u$
component is essentially the same, according to Eq. (\protect\ref{dP/dz2}). }
\label{fig:l_amp_gamma_0.1}
\end{figure}

\begin{figure}[tbp]
\centering
\subfigure[]{
\includegraphics[width=0.5\textwidth]{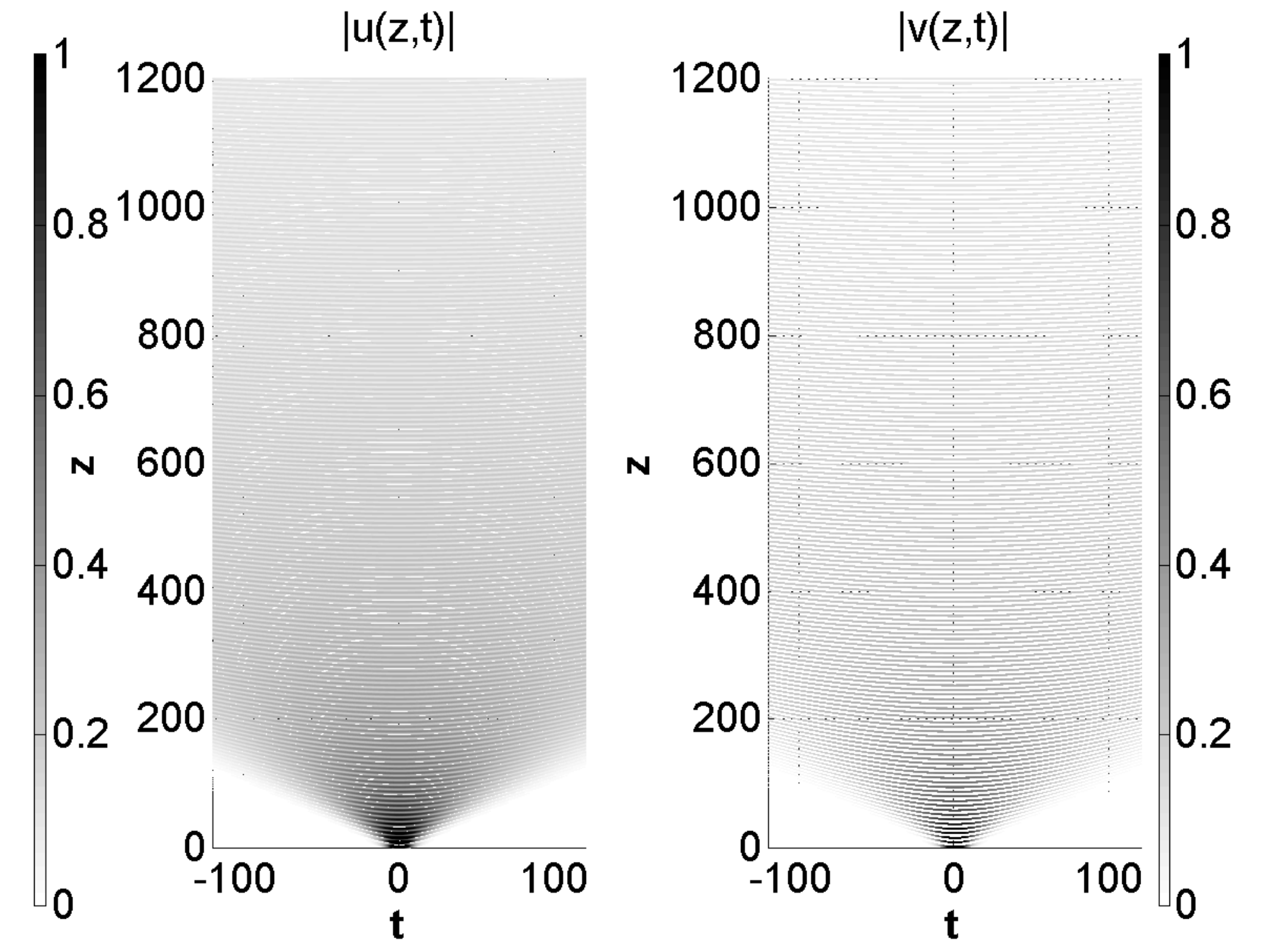}}
\subfigure[]{
\includegraphics[width=0.5\textwidth]{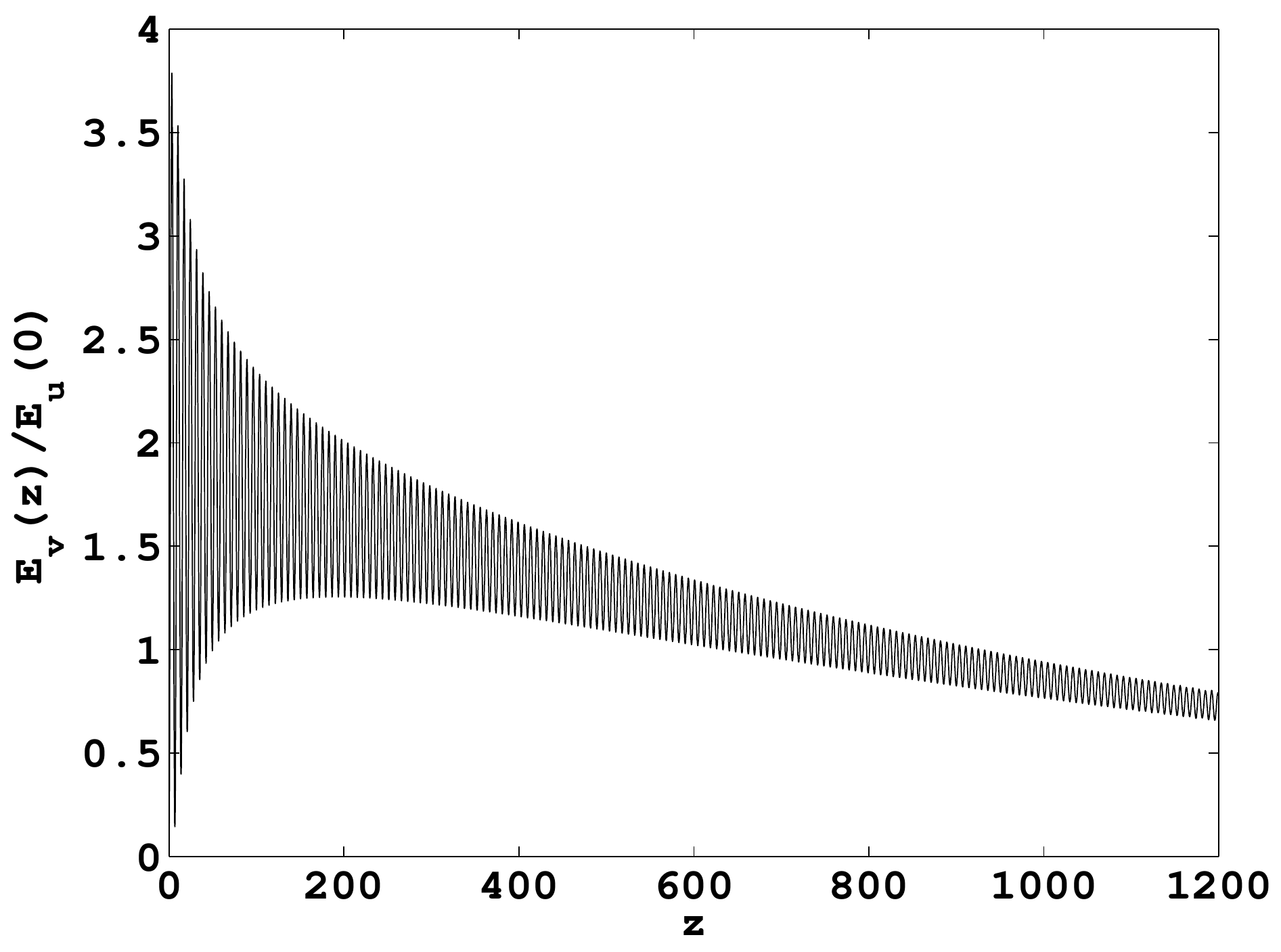}}
\caption{The same as in Fig. \protect\ref{fig:l_amp_gamma_0.1}, but for the
strong coupling, $\protect\gamma =0.9$. }
\label{fig:l_amp_gamma_0.9}
\end{figure}

Next, we consider the situation close to the $\mathcal{CPT}$%
-symmetry-breaking threshold (\ref{thr}), namely, with $\gamma =0.9$ for $%
q=1 $. The respective numerical solution, generated by initial conditions (%
\ref{IC1}), is displayed in Fig. \ref{fig:l_amp_gamma_0.9}, which,
naturally, demonstrates strong coupling between the two components and more
dramatic evolution. In this case too, the asymptotic stage of the evolution
for broad pulses is correctly predicted by the above-mentioned analytical
approximation.

\begin{figure}[h]
\centering
\subfigure[]{
\includegraphics[width=0.5\textwidth]{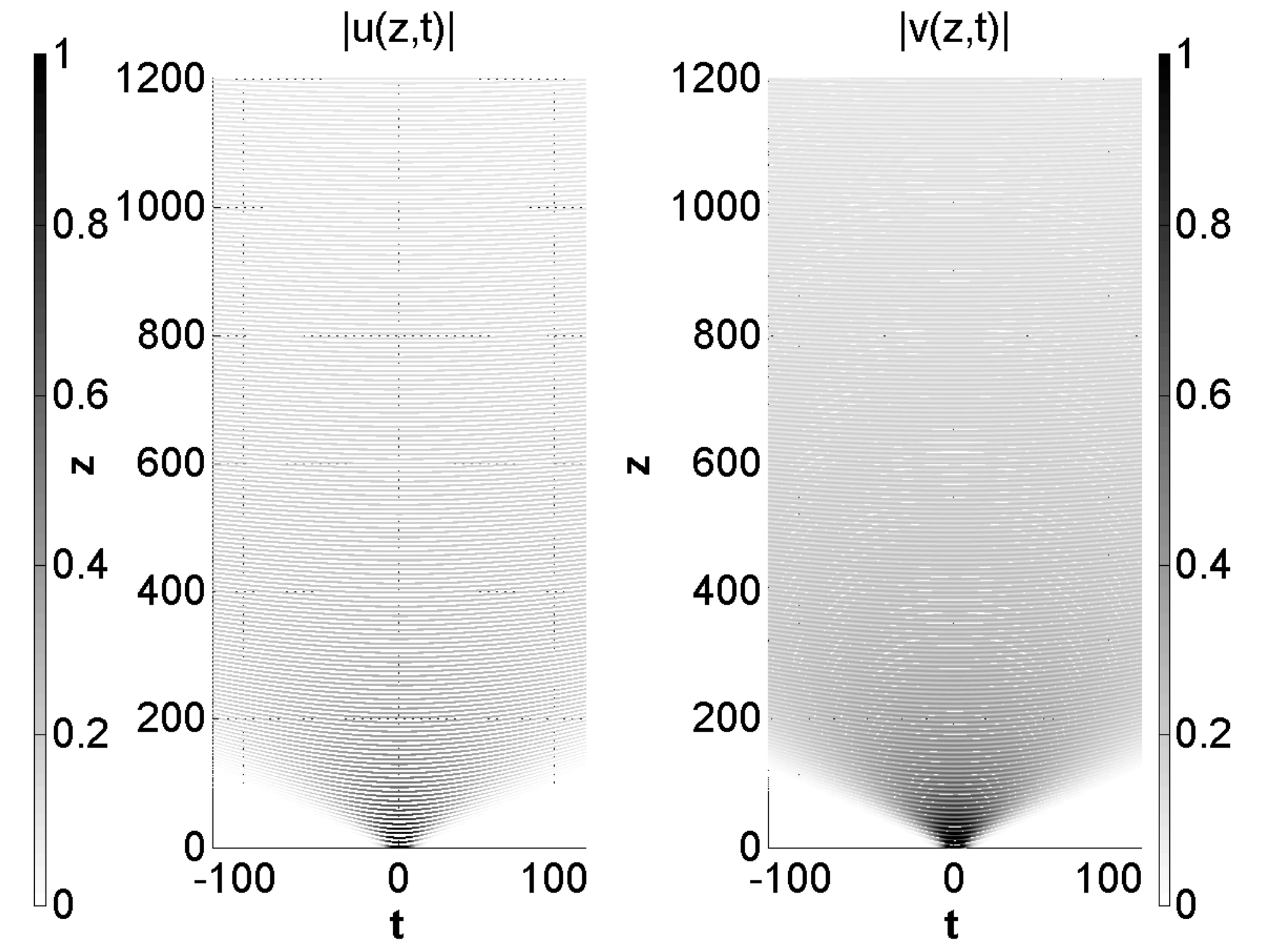}}
\subfigure[]{
\includegraphics[width=0.5\textwidth]{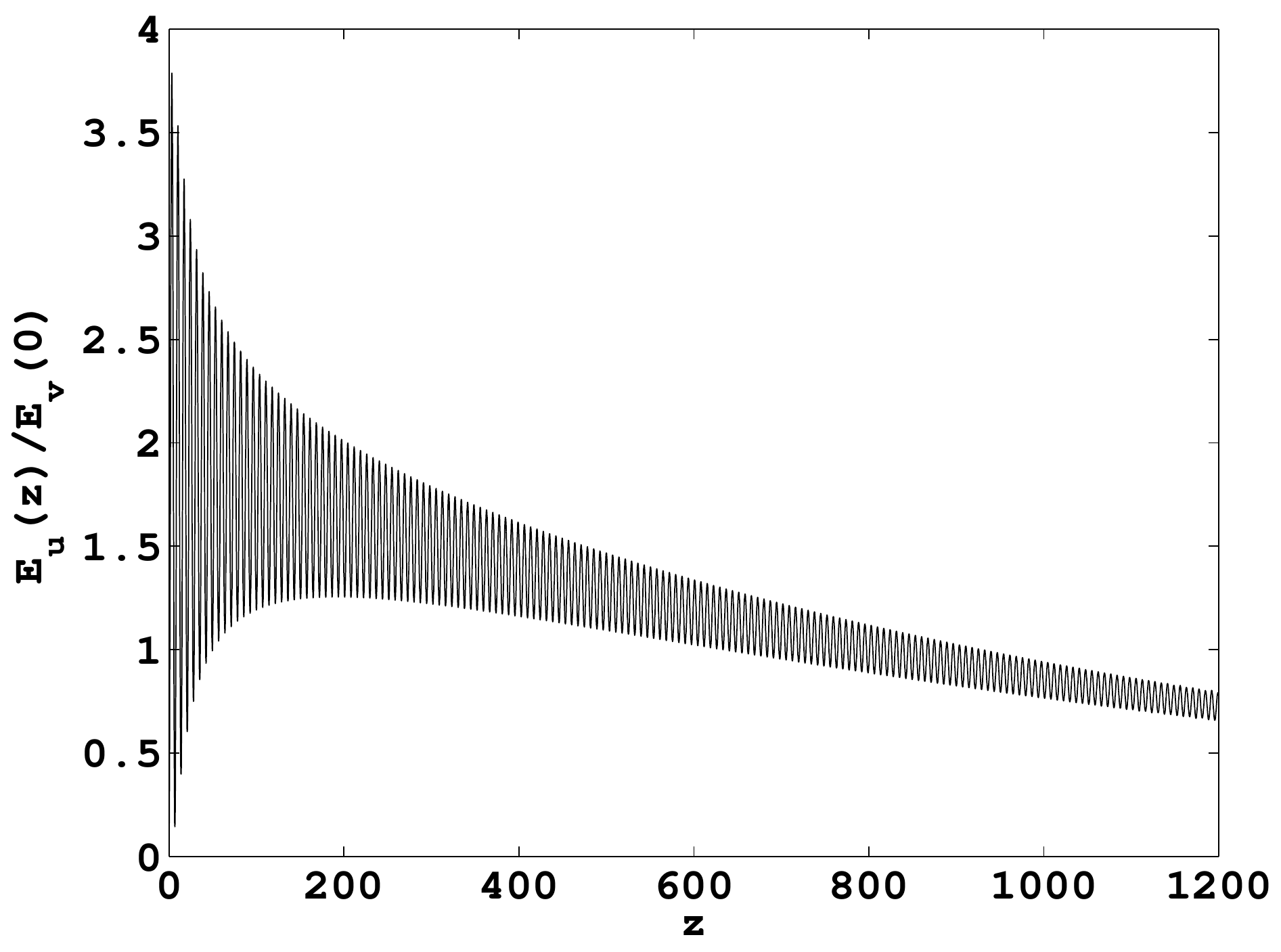}}
\caption{The same as in Fig. \protect\ref{fig:l_amp_gamma_0.9}, but for
initial conditions (\protect\ref{IC2}), with the difference that panel (b)
displays the evolution of the energy in the $u$ component.}
\label{fig:v_amp_gamma_0.9}
\end{figure}

To test the symmetry of the system, we have also performed simulations of
the evolution starting from initial conditions (\ref{IC2}), with swapped
components $u$ and $v$. Comparison of the respective results, displayed in
Figs. \ref{fig:v_amp_gamma_0.9}(a) and \ref{fig:v_amp_gamma_0.9}(b), with
their counterparts shown above in Fig. \ref{fig:l_amp_gamma_0.9} confirms
the symmetry. Furthermore, the detailed comparison of the real and imaginary
parts of the two components in both cases (not shown here in detail) exactly
corroborates the full $\mathcal{CP}$ symmetry implied by definition (\ref%
{CPT}).

For $\gamma >q$, when the the $\mathcal{CP}$ symmetry of the system is
broken, according to Eq. (\ref{thr}), direct simulations (not shown here)
demonstrate blowup of solutions, as should be expected above the
symmetry-breaking point \cite{Ham,nature}.

\begin{figure}[h]
\centering
\subfigure[]{
\includegraphics[width=0.6\textwidth]{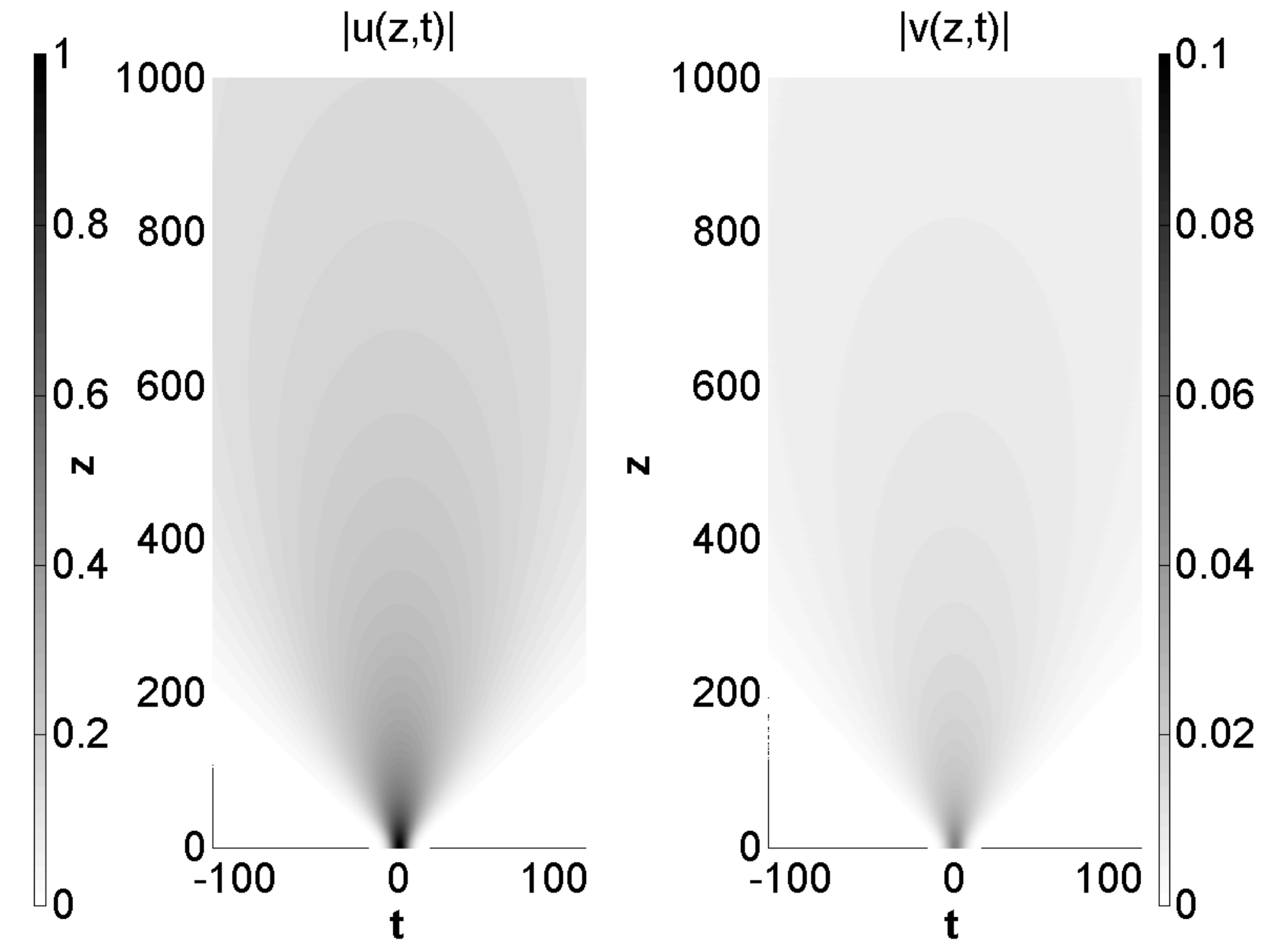}}
\subfigure[]{
\includegraphics[width=0.6\textwidth]{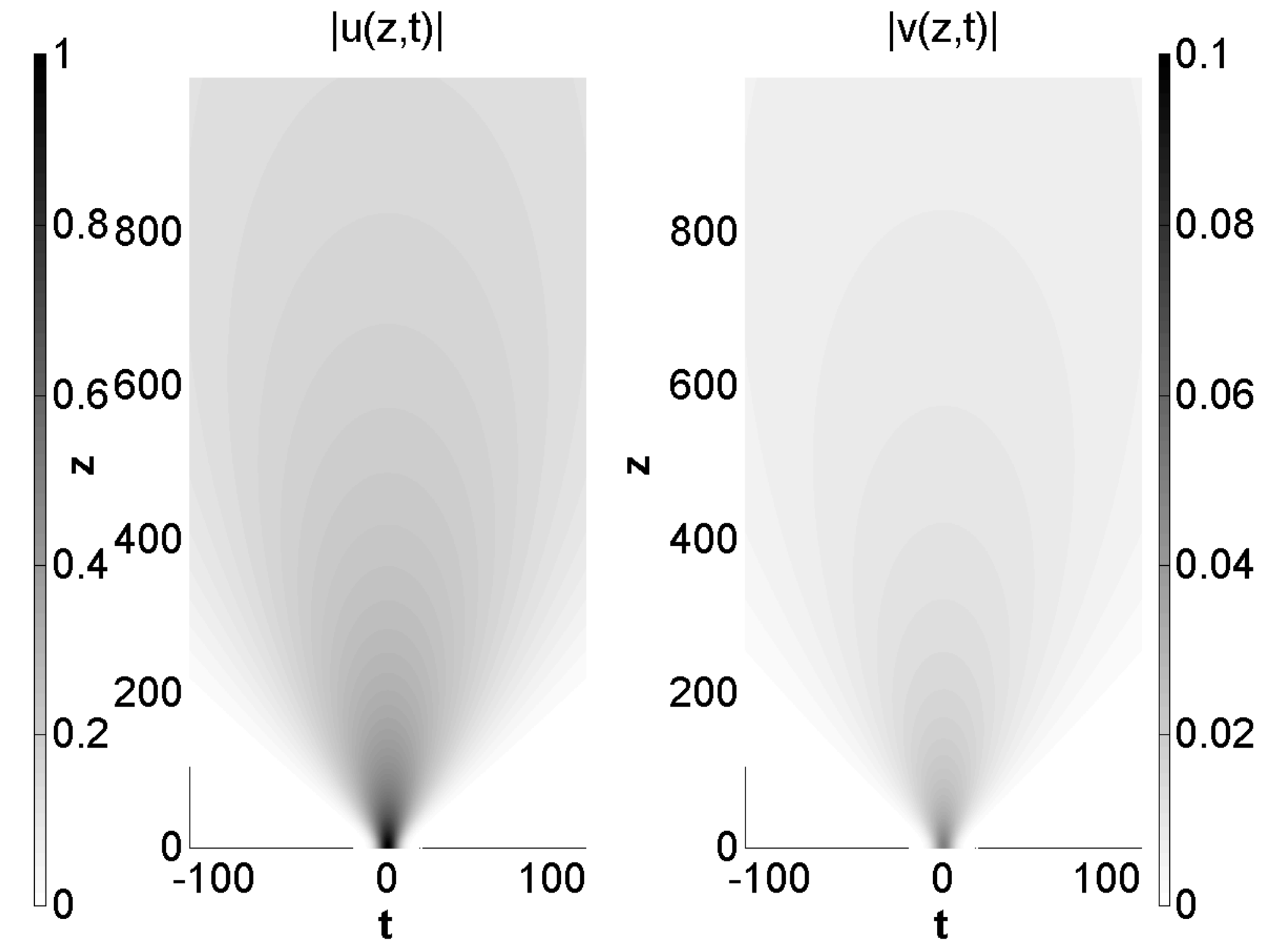}}
\caption{(a) The numerical simulations for absolute values $|u(z,t)|$ and $%
|v(z,t)|$ in the weak-coupling regime, with $\protect\gamma =0.1$, $q=1$ and
the initial conditions taken as per Eqs. (\protect\ref{v-u}) and (\protect
\ref{Gauss}) (with the upper sign) at $z=0$. (b) The respective analytical
solutions.}
\label{fig:u_numeric_gamma_0.1}
\end{figure}

The analytical approximation for broad pulses, based on Eqs. (\ref{v-u}) and
({\ref{Gauss})}, was directly tested by comparing its predictions with the
numerically simulated evolution commencing from the initial conditions
produced by Eqs. (\ref{v-u}) and ({\ref{Gauss}) with $z=0$ and the upper
sign in the latter equation. Figure \ref{fig:u_numeric_gamma_0.1} shows that
the respective analytical and the numerical results are almost identical.
The comparison produces equally good results (not shown here in detail) if
the initial conditions are taken, instead, as per }Eq. (\ref{u-v}) and Eq. ({%
\ref{Gauss}) with the lower sign, at }$z=0${. }

On the other hand, for strong coupling, e.g., for $\gamma =0.9$, when $\exp
\left( \pm i\sqrt{q^{2}-\gamma ^{2}}z\right) $ is no longer a rapidly
oscillating carrier in comparison with slowly varying $\tilde{u}$ and $%
\tilde{v}$, see Eqs. (\ref{v-u}) and (\ref{u-v}), the analytical
approximation is no longer relevant. The comparison with the numerical
results corroborates this expectation (not shown here in detail either).

\subsection{The nonlinear system}

Simulations of the nonlinear system, based on Eqs. (\ref{u}) and (\ref{v}),
were performed by varying the nonlinearity coefficient, $\sigma $, and (as
above) the coupling coefficient, $\gamma $. The initial conditions were
taken in the form of Eq. (\ref{IC1}), unless stated otherwise.

\begin{figure}[tbp]
\centering
\subfigure[]{
\includegraphics[width=0.5\textwidth]{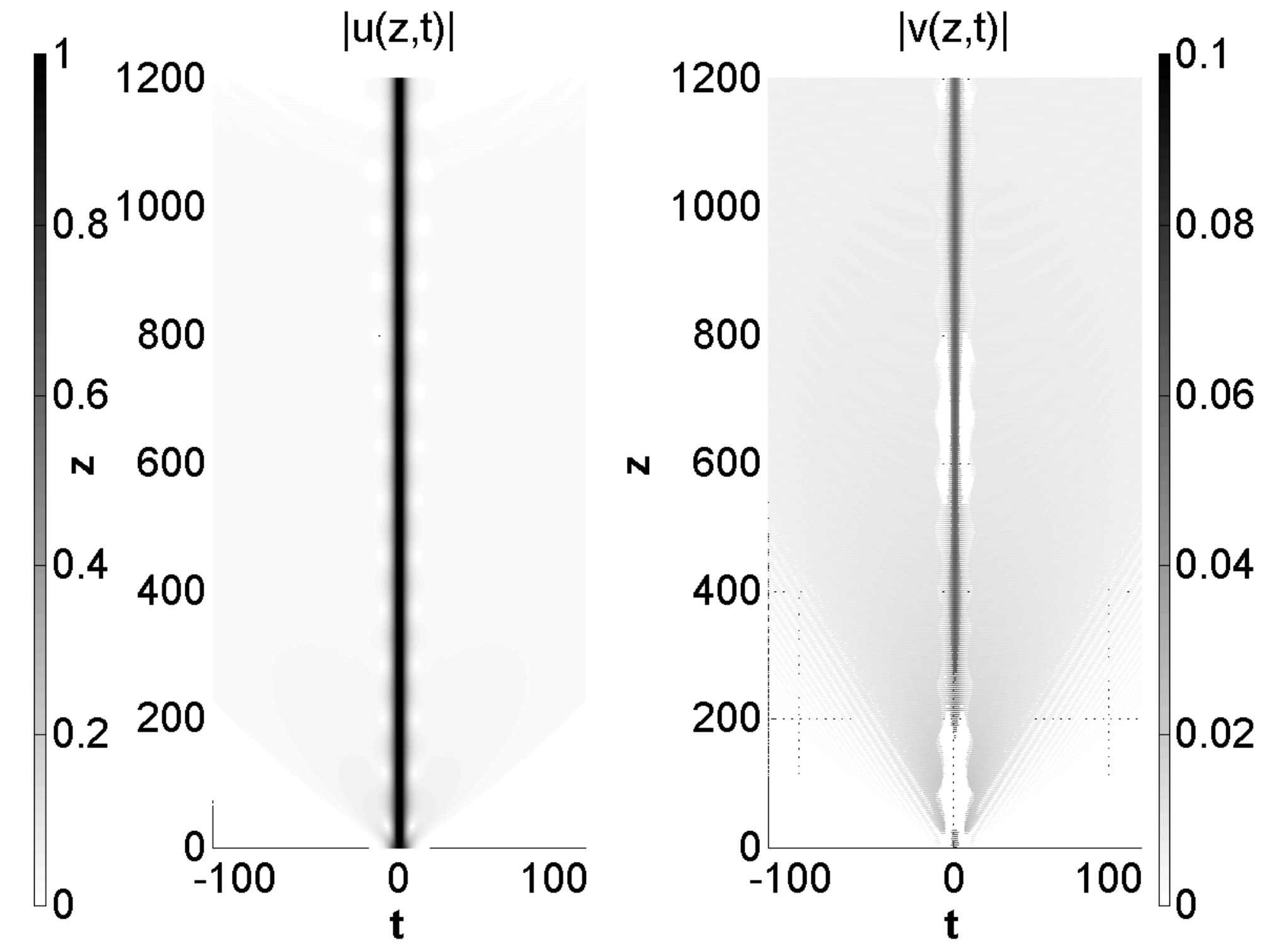}}
\subfigure[]{
\includegraphics[width=0.5\textwidth]{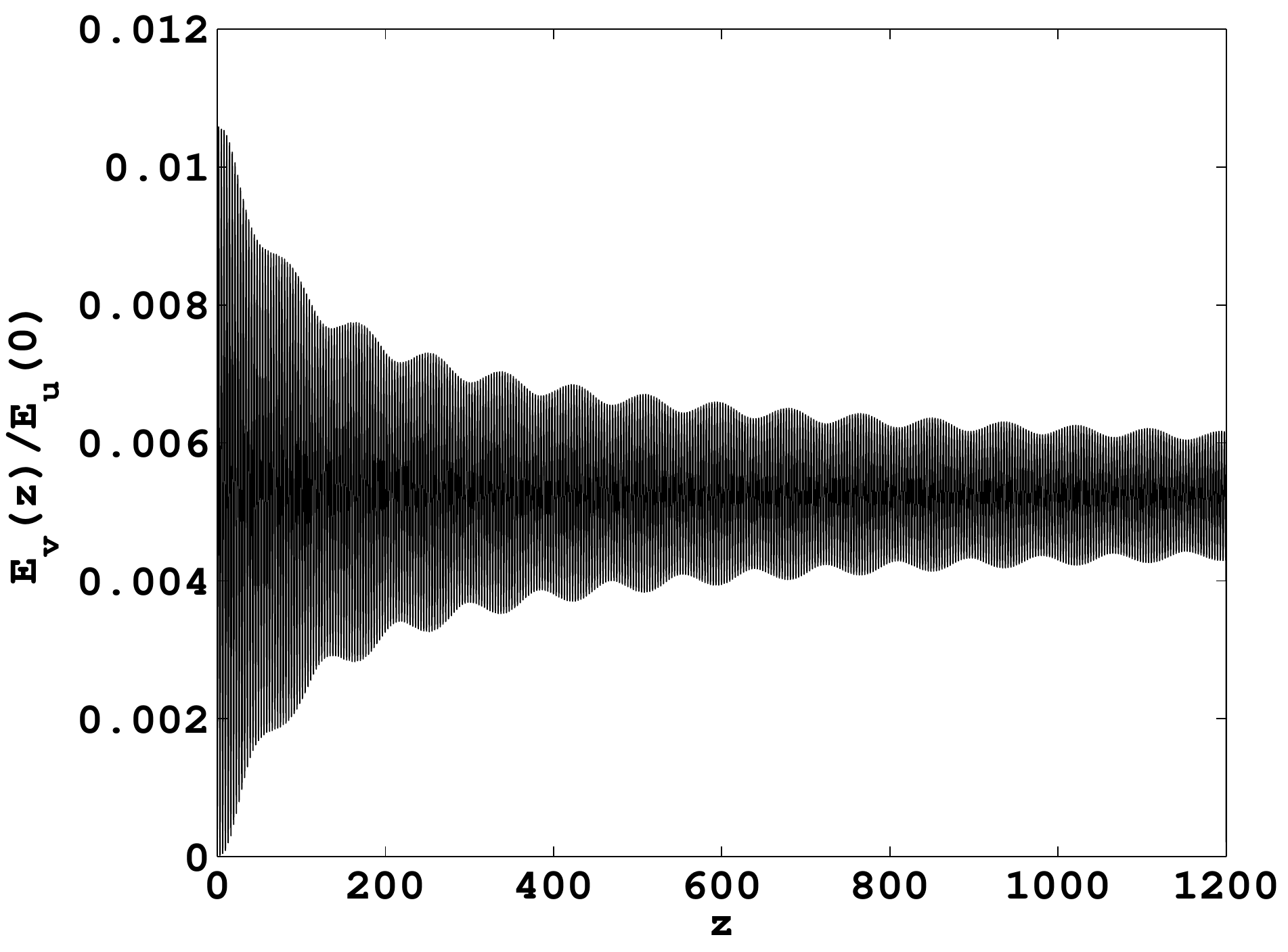}}
\caption{The same as in Fig. \protect\ref{fig:l_amp_gamma_0.1}, but produced
by simulations of the full nonlinear system (\protect\ref{u}), (\protect\ref%
{v}) with $q=1$, $\protect\gamma =0.1$, and $\protect\sigma =0.1$. The weak
radiation field around the emerging quasi-soliton is virtually invisible, if
local powers, $|u(z,t)|^{2}$ and $|v(z,t)|^{2}$, are displayed instead of
the amplitudes, $|u(z,t)|$ and $|v(z,t)|$.}
\label{fig:amp_nl_0.1_gamma_0.1}
\end{figure}

We start by addressing the weakly coupled system with weak nonlinearity,
\textit{viz}., the one with $0<\sigma \ll 1$ and $0<\gamma \ll q\equiv 1$.
For $\gamma =0.1$ and $\sigma =0.1$, Figs. \ref{fig:amp_nl_0.1_gamma_0.1}(a)
and (b) demonstrate that the focusing nonlinearity readily causes
self-trapping of a robust oscillating quasi-soliton. Thus, the weak
nonlinearity, while breaking the $\mathcal{CP}$ symmetry (see above),
creates the self-confined modes.

Next, we increase the strength of the coupling to $\gamma =0.8$, keeping the
nonlinear term small, with $\sigma =0.1$. In this case, Figs. \ref%
{fig:amp_nl_0.1_gamma_0.8}(a) and (b) demonstrate strong self-focusing of
the modes occurring around $z=10$, followed by the propagation of the
confined mode in a sufficiently robust form, although with more conspicuous
emission of radiation waves than in the case of $\gamma =0.1$, cf. Fig. \ref%
{fig:amp_nl_0.1_gamma_0.1}(a). Thus, in this case too, the system tends to
form oscillatory quasi-soliton modes.

The formation of these solitons is readily explained by Eq. (\ref{slow}) for
$\tilde{u}$. Indeed, it is easy to check that the width and amplitude of the
emerging solitons satisfy conditions (\ref{broad}). The solitons are
observed in Figs. \ref{fig:amp_nl_0.1_gamma_0.1} and \ref%
{fig:amp_nl_0.1_gamma_0.8} in an oscillatory form, which is different from
the stationary solution (\ref{sol}), in accordance with the well-knows fact
that perturbed NLS solitons may feature long-lived vibrations, similar to
those observed in these figures \cite{Anderson1,Anderson2}.
\begin{figure}[tbp]
\centering
\subfigure[]{
\includegraphics[width=0.5\textwidth]{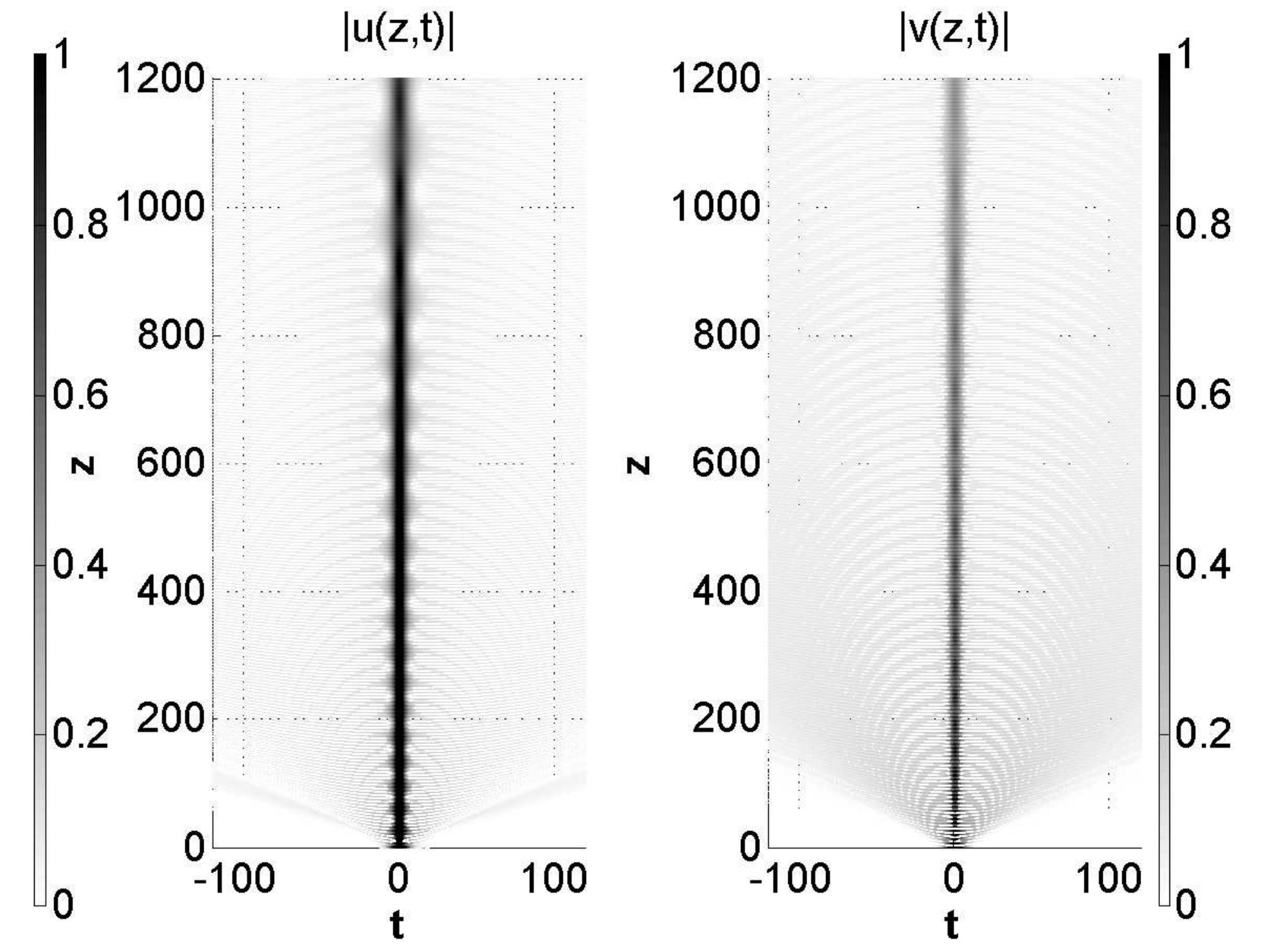}}
\subfigure[]{
\includegraphics[width=0.5\textwidth]{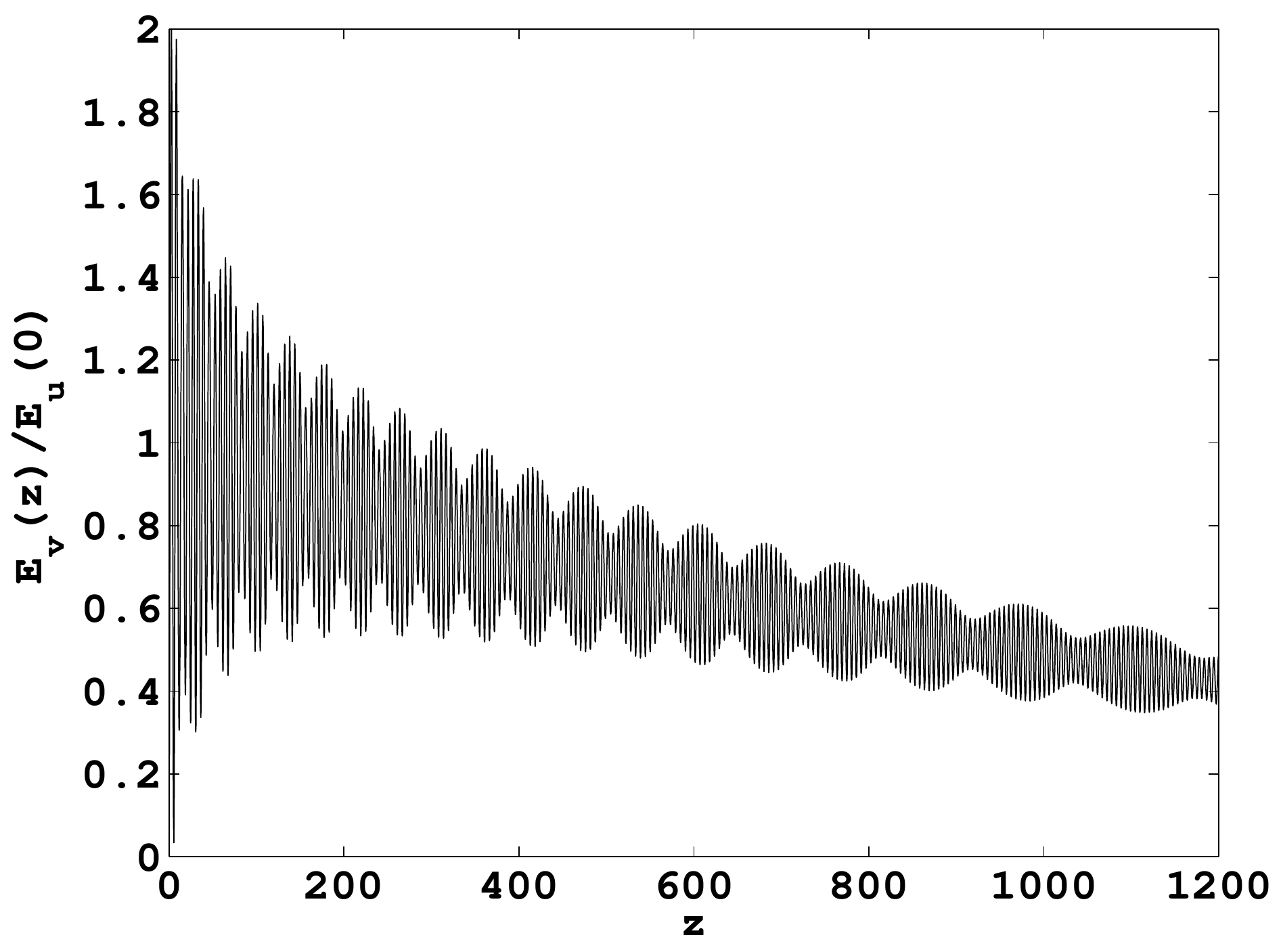}}
\caption{The same as in Fig. \protect\ref{fig:amp_nl_0.1_gamma_0.1}, but in
the case of strong coupling, $\protect\gamma =0.8$.}
\label{fig:amp_nl_0.1_gamma_0.8}
\end{figure}

For $\gamma \geq 0.9$ and the same weak nonlinearity, $\sigma =0.1$, the
simulations demonstrate that amplitudes of both modes, $u$ and $v$, diverge
after a short propagation distance, which implies that the symmetry breaking
takes place in these cases, that are close to the threshold (\ref{thr}) of
the symmetry breaking. The small difference of $\gamma =0.9$ from the exact
linear threshold, $\gamma =1$, is compensated in this case by the presence
of the nonlinearity which, as said above, is also a $\mathcal{CP}$%
-symmetry-breaking factor.

The swap of initial conditions (\ref{IC1}) and (\ref{IC2}) in the nonlinear
system produces a strong effect. Indeed, in the above simulations, performed
for input (\ref{IC1}), the pulse was launched in component $u$, where the
nonlinearity is self-focusing [see Eq. (\ref{u})], while initial conditions (%
\ref{IC2}) imply that the pulse is launched into component $v$ with the
self-defocusing cubic term, see Eq. (\ref{v}). Accordingly, in the latter
case, the simulations produce the results displayed in Fig.\ \ref%
{missing_v_amp_nl_0.1_gamma_0.8}: instead of the quick self-trapping (cf.
Fig. \ref{fig:amp_nl_0.1_gamma_0.8}), the pulse launched in the $v$
component features slow expansion. An additional difference is that the
frequency of oscillations observed in the latter case is approximately half
of that observed in Fig. \ref{fig:amp_nl_0.1_gamma_0.8}.
\begin{figure}[tbp]
\centering
\subfigure[]{
\includegraphics[width=0.5\textwidth]{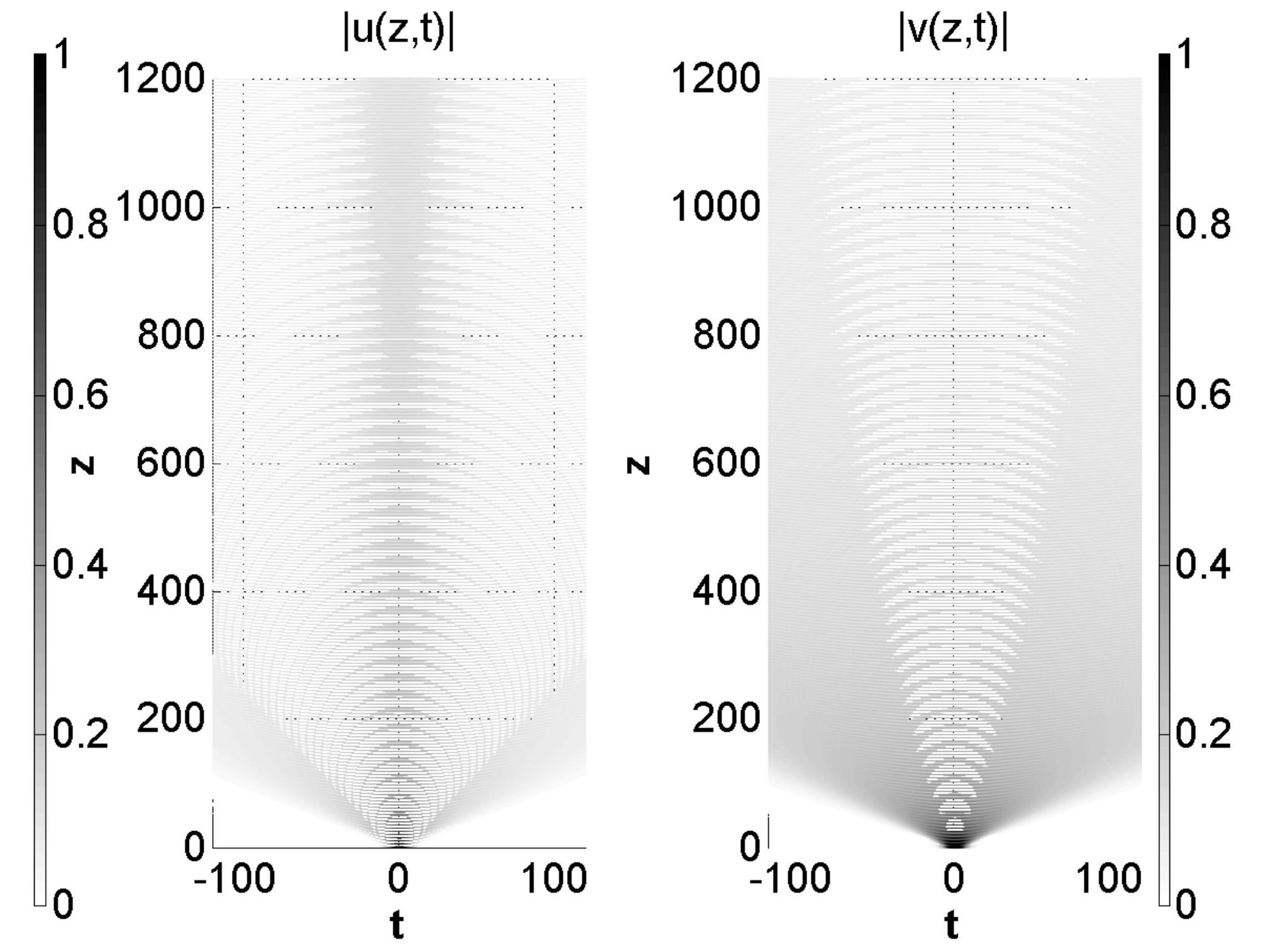}}
\subfigure[]{
\includegraphics[width=0.5\textwidth]{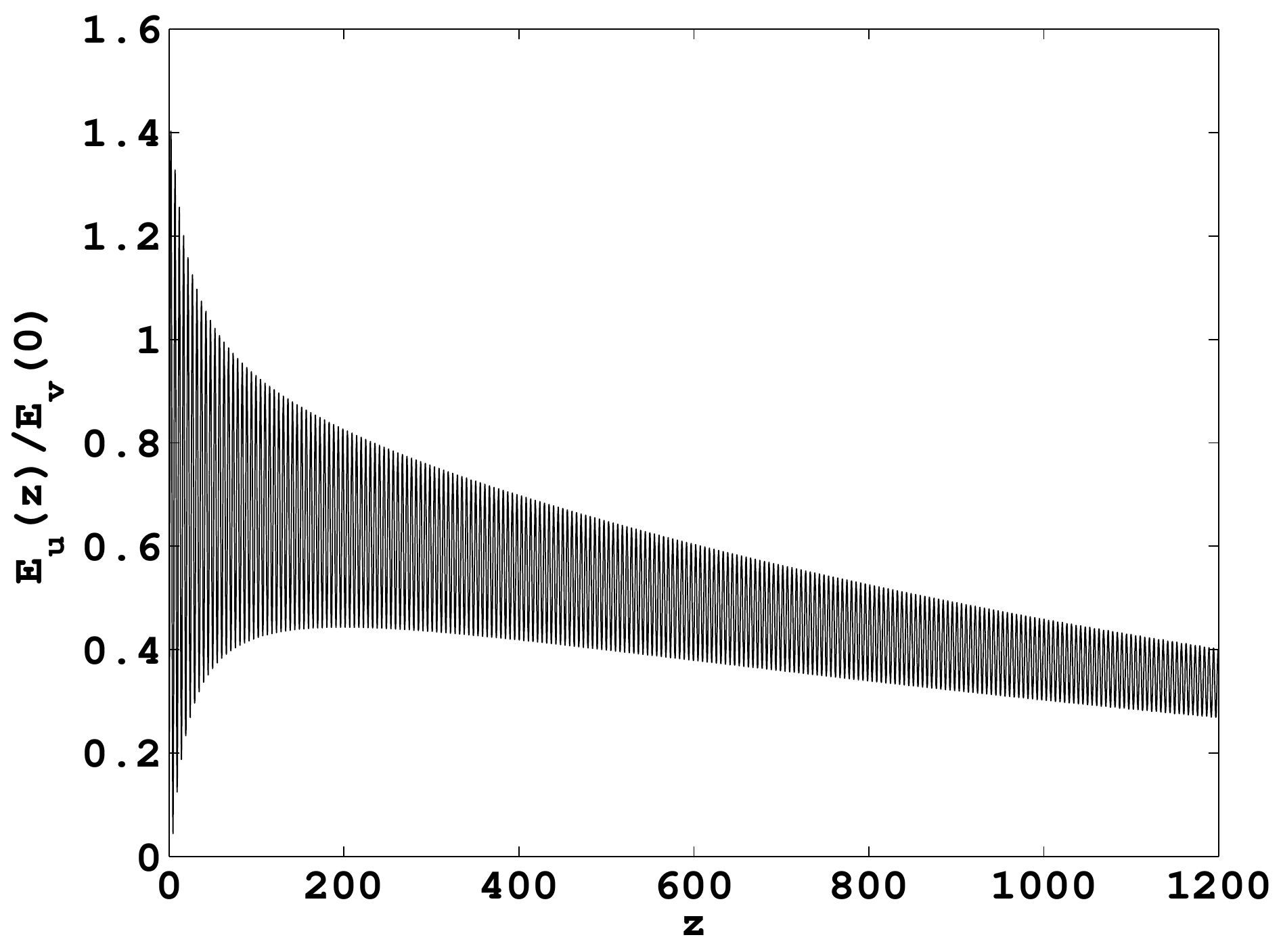}}
\caption{The same as in Fig. \protect\ref{fig:amp_nl_0.1_gamma_0.8}, but for
initial conditions (\protect\ref{IC2}).}
\label{missing_v_amp_nl_0.1_gamma_0.8}
\end{figure}

The increase of $\sigma $ at a fixed coupling constant, $\gamma $, enhances
the $\mathcal{CP}$-symmetry-breaking effects, and eventually leads to
destruction of the quasi-soliton. In particular, for the weakly coupled
system, with $\gamma =0.1$, the destabilization of the quasi-soliton sets in
at critical value $\sigma =0.5$, as shown in Fig. \ref%
{fig:amp_nl_0.1_gamma_0.1}. In this case, the integral energy slowly grows
with $z$, which is followed by blowup at very large values of $z$ (not shown
here in detail).
\begin{figure}[tbp]
\centering
\includegraphics[width=0.5\textwidth]{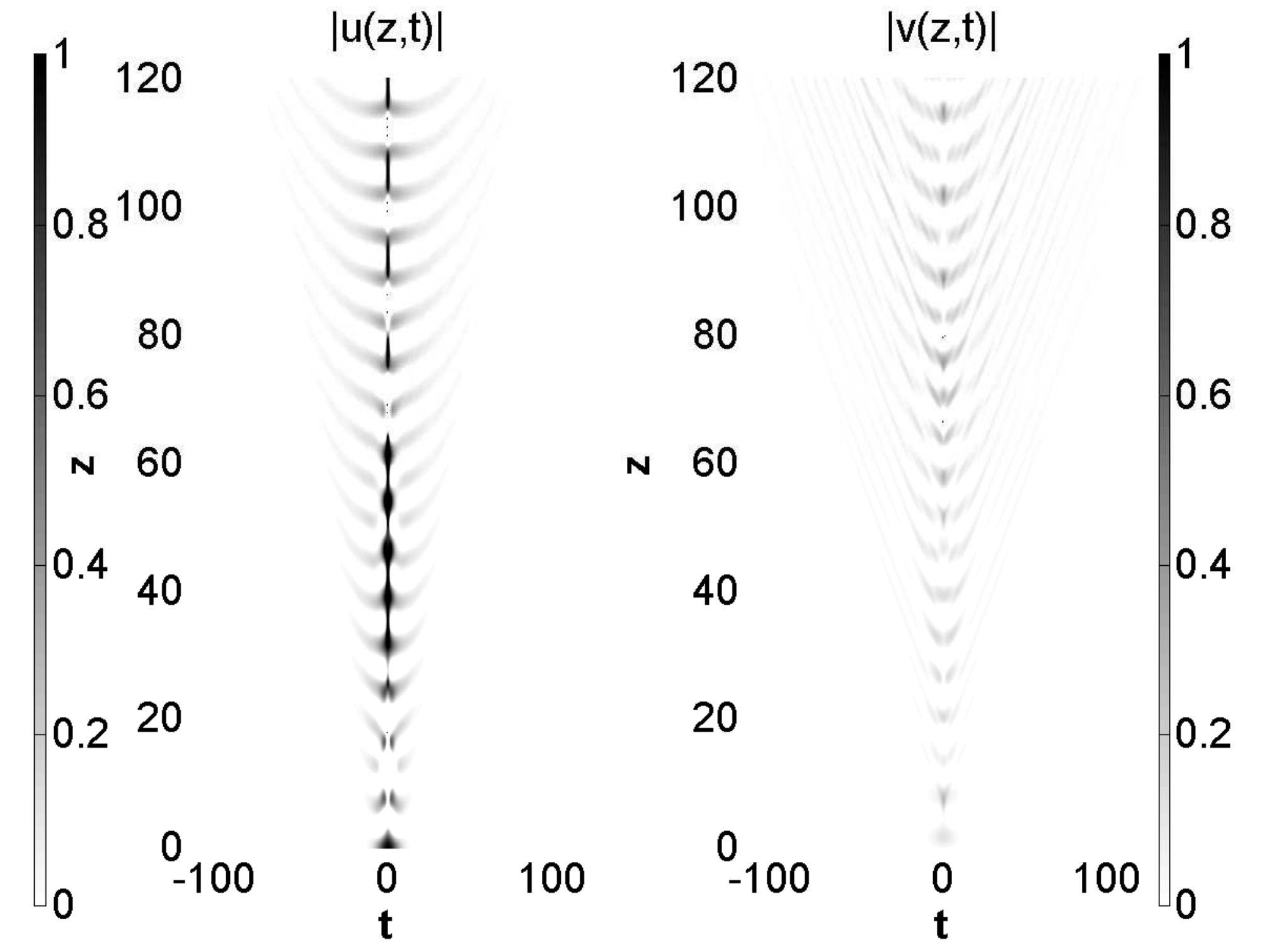}
\caption{The same as in Fig. \protect\ref{fig:amp_nl_0.1_gamma_0.1}(a), but
at the critical value of the nonlinearity strength, $\protect\sigma =0.5$,
at which the gradual destruction of the quasi-soliton commences.}
\label{fig:amp_nl_0.5_gamma_0.1}
\end{figure}

\begin{figure}[h]
\centering
\includegraphics[width=0.5\textwidth]{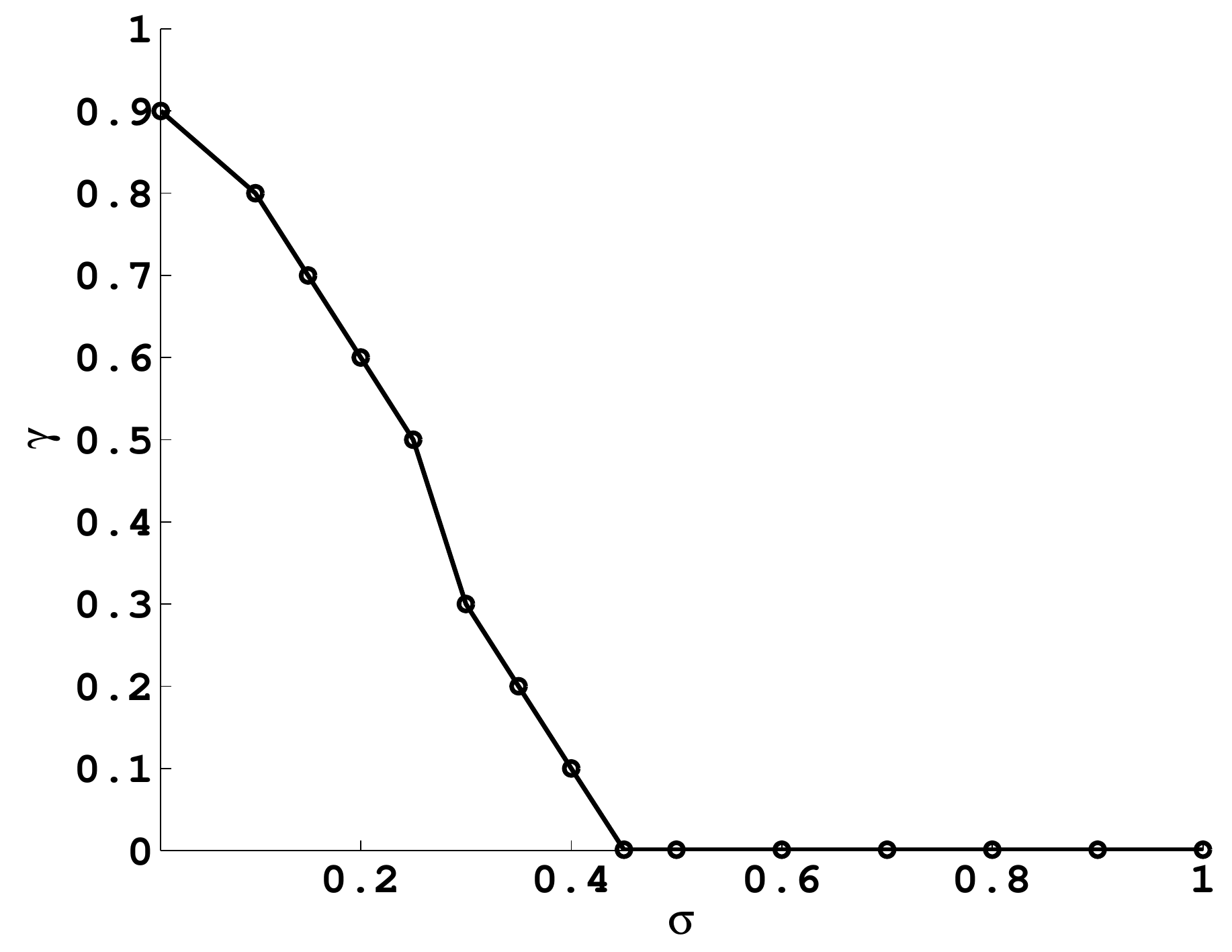} 
\caption{The stability region (beneath the curve) for the oscillatory
quasi-solitons created from initial conditions (\protect\ref{IC1}), in the
plane of $\left( \protect\sigma ,\protect\gamma \right) $. }
\label{fig:stable_nl_vs_gamma}
\end{figure}

By means of systematical simulations, we have collected the critical values
of $\sigma $, at which the quasi-soliton suffers the onset of the
destabilization, eventually leading to the blowup, at increasing values of
the coupling constant, $\gamma $. The corresponding dependence between $%
\sigma $ and $\gamma $, shown in Fig. (\ref{fig:stable_nl_vs_gamma}),
naturally demonstrates that the critical strength of the nonlinearity
vanishes when $\gamma $ approaches the threshold of the symmetry breaking in
the linear system, $\gamma =1$, see Eq. (\ref{thr}) [recall the
normalization is fixed by setting $q=1$ in Eqs. (\ref{u}) and (\ref{v})].

\begin{figure}[tbp]
\centering
\subfigure[]{
\includegraphics[width=0.6\textwidth]{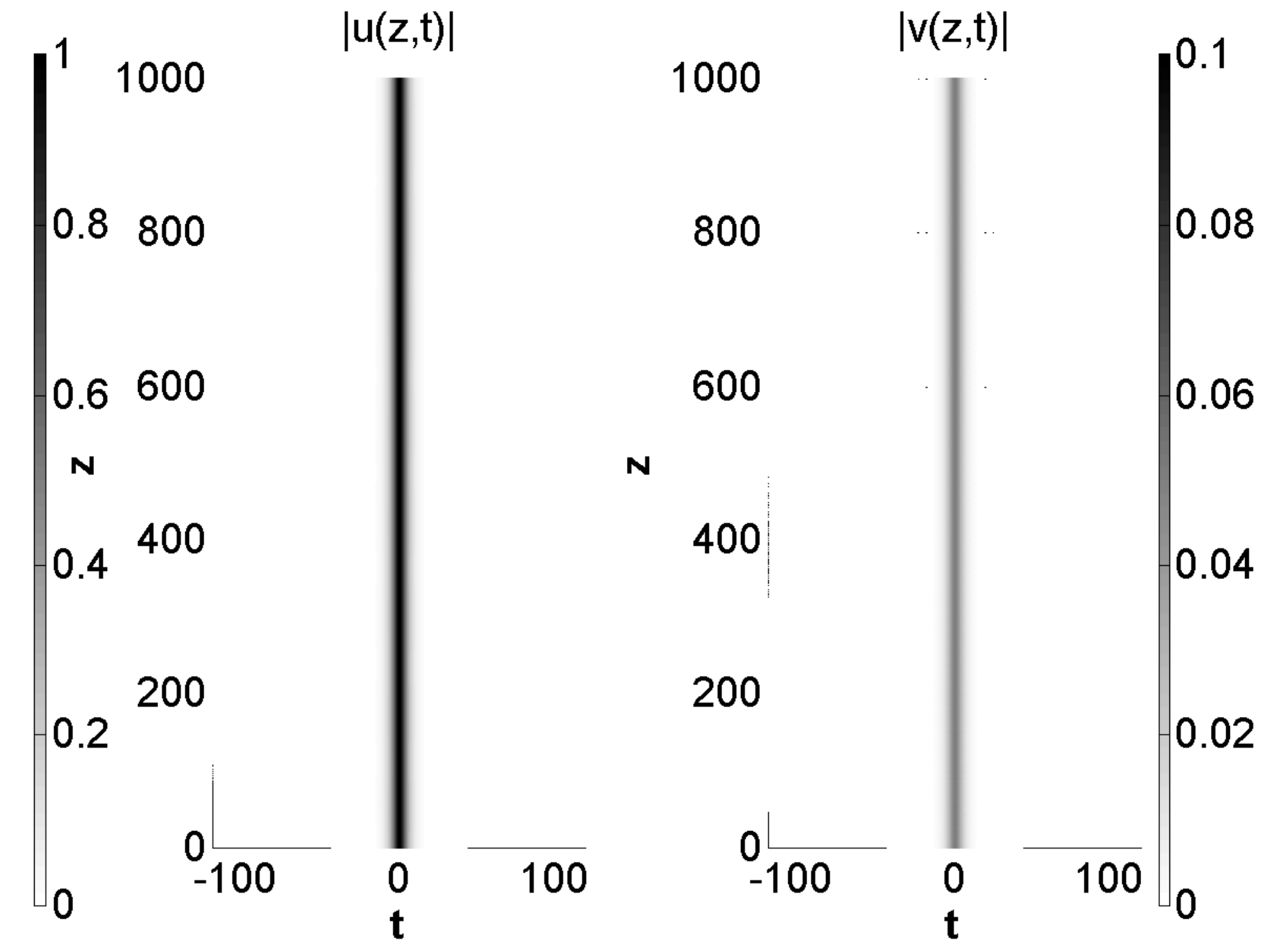}}
\subfigure[]{
\includegraphics[width=0.6\textwidth]{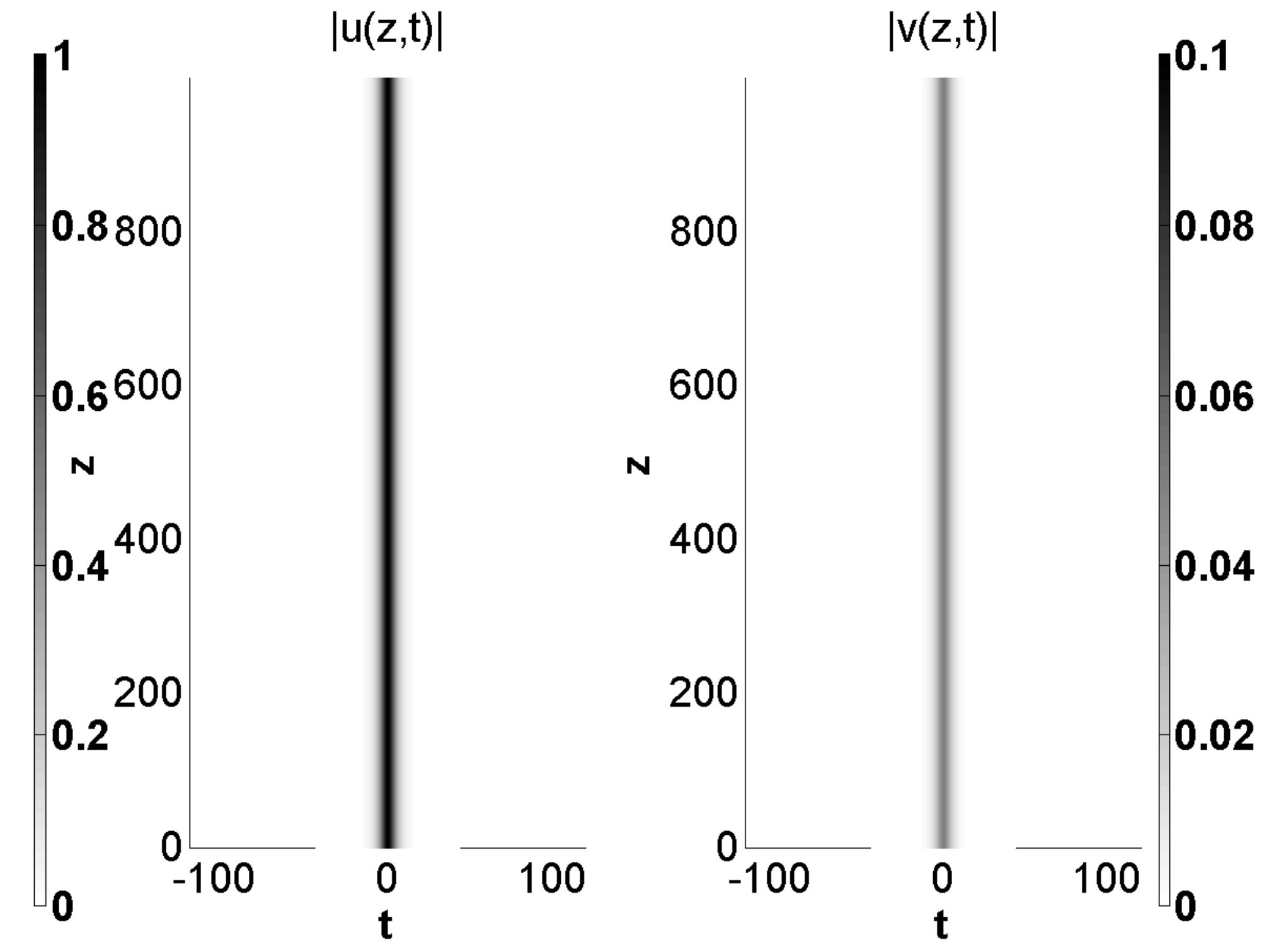}}
\caption{(a) The numerical simulations for absolute values $|u(z,t)|$ and $%
|v(z,t)|$ in the weakly coupled nonlinear system, with $\protect\gamma =0.1$%
, $\protect\sigma =0.1$, $q=1$ and initial conditions taken as per Eqs. (%
\protect\ref{v-u}) and (\protect\ref{sol}) with $\protect\kappa =0.05$ at $z=0$. (b) The respective analytical solution.}
\label{fig:u_numeric_gamma_0.1_nl}
\end{figure}

The analytical approximation based on Eqs. (\ref{v-u}) and ({\ref{sol})} was
tested for the broad solitons too{. Figure \ref{fig:u_numeric_gamma_0.1_nl}
shows that the respective analytical and the numerical results are almost
identical, thus validating the analytical approximation for the nonlinear
system. }

\subsection{Stationary gap solitons in the nonlinear system}

The quasi-solitons considered above are built as breathers, featuring
permanent oscillations in both components. On the other hand, Eqs. (\ref{0}-%
\ref{Vexact}) predict the existence of stationary gap solitons in the same
system. To check this possibility in the numerical form, we solved Eqs. (\ref%
{U}-\ref{V}) by means of the Newton's method \cite{davis1984numerical}. This
was done using the approximate analytical solution, given by Eqs. (\ref{0})
and (\ref{Vexact}), as the initial guess. The results are produced here for $%
\sigma =1$ and different values of the coupling constant, $\gamma $.

In the case of weak coupling, $\gamma =0.1$, when it is natural to expect
the solutions to be strongly asymmetric, in terms of the two components, the
numerical solution at $k=0$, i.e., at the center of the bandgap, is very
close to its above-mentioned analytical counterpart, as seen in Fig. \ref%
{fig:c_gamma_0.1}.
\begin{figure}[tbp]
\centering
\includegraphics[width=0.5\textwidth]{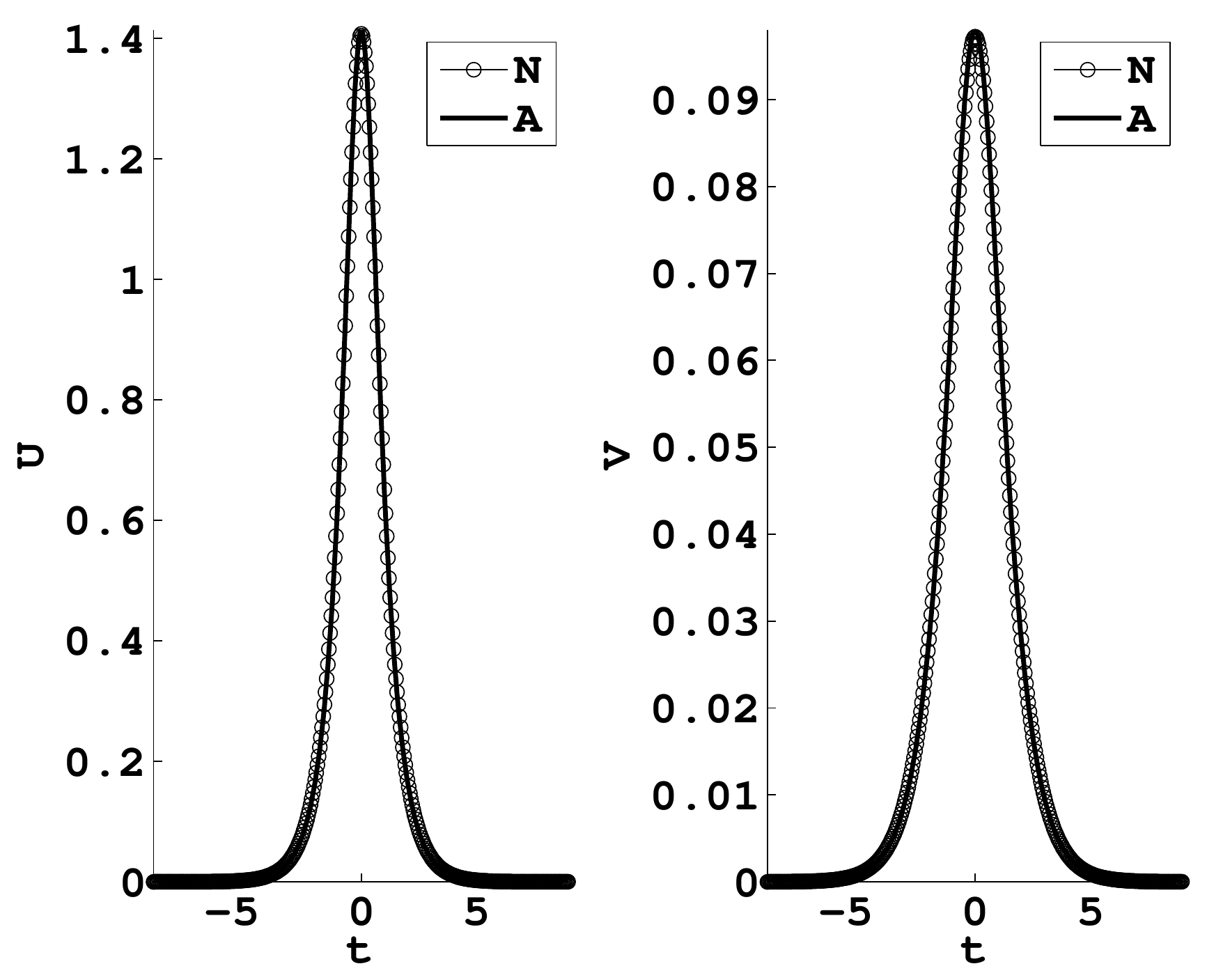}
\caption{Numerical (\textquotedblleft Newton" - N) and analytical solutions (A) for
the two components of a strongly asymmetric gap soliton, $U(t)$ and $V(t)$
(note different scales in the panels), obtained from Eqs. (\protect\ref{U})
and (\protect\ref{V}) for $q=1$, $\protect\sigma =1$, $\protect\gamma =0.1$,
and $k=0$. The respective analytical solution is given by Eq. (\protect\ref%
{0}) and (\protect\ref{Vexact}). }
\label{fig:c_gamma_0.1}
\end{figure}
For larger values of $\gamma $, the numerical solution differs from the
analytical one, obtained under condition $\gamma \ll q\equiv 1$, although
the difference remains relatively small for $\gamma =0.5$, as shown in Fig. %
\ref{fig:c_gamma_0.5}. The difference becomes essential for $\gamma =0.9$
[in fact, very close to the symmetry-breaking threshold (\ref{thr})], as can
be seen in Fig. \ref{fig:c_gamma_0.9}.
\begin{figure}[tbp]
\centering
\includegraphics[width=0.5\textwidth]{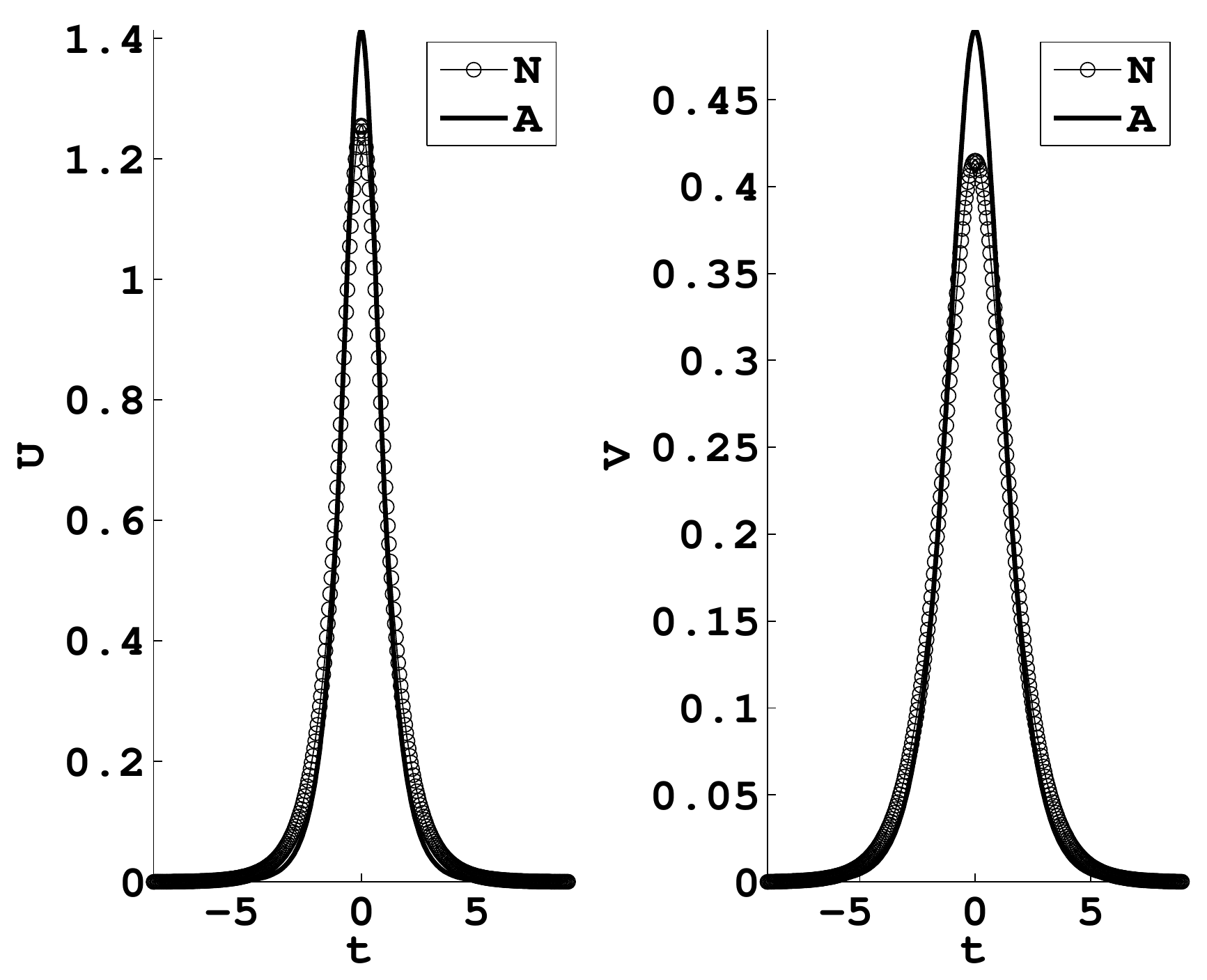}
\caption{The same as in Fig. \protect\ref{fig:c_gamma_0.1}, but for $\protect%
\gamma =0.5$.}
\label{fig:c_gamma_0.5}
\end{figure}
\begin{figure}[tbp]
\centering
\includegraphics[width=0.5\textwidth]{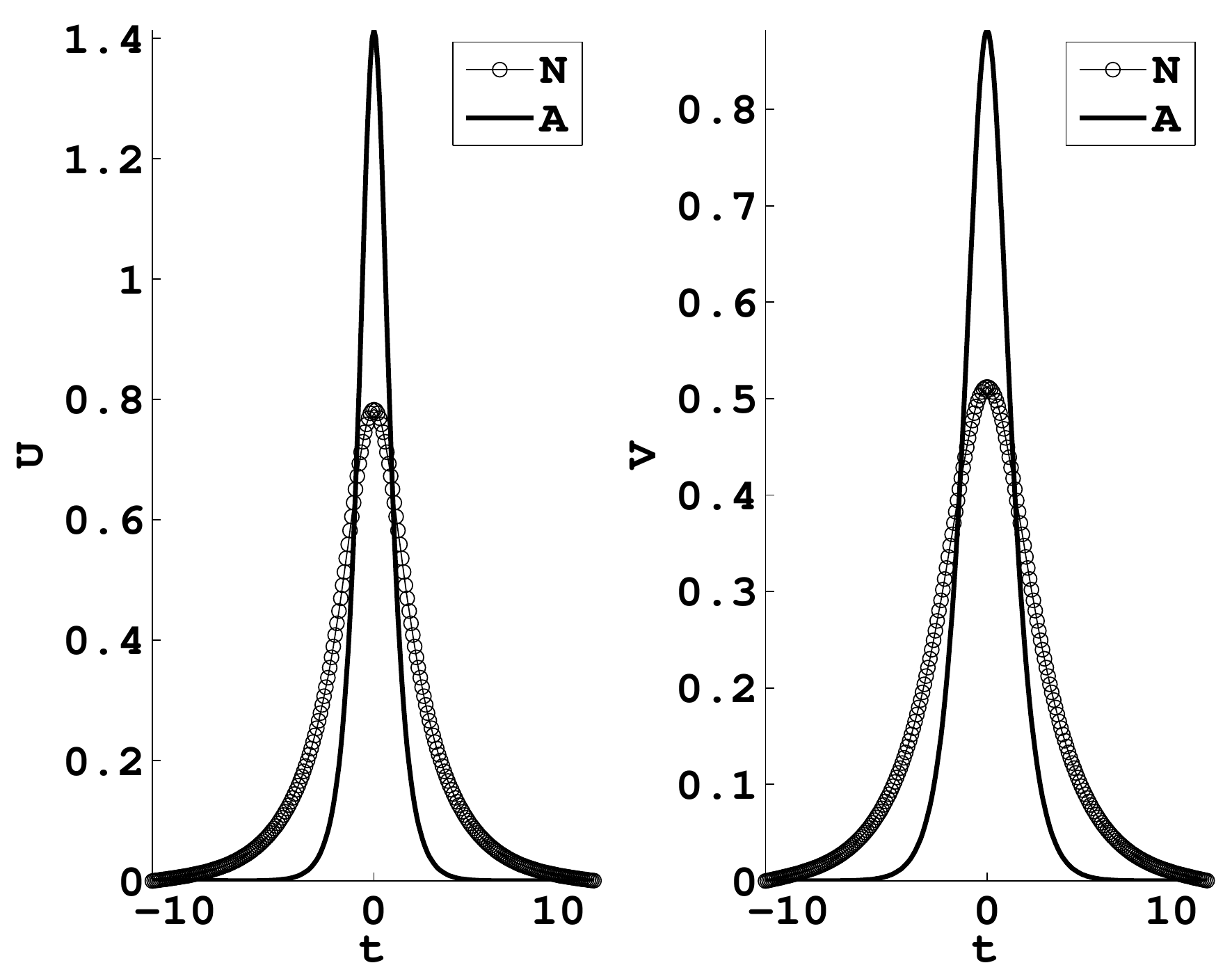}
\caption{The same as in Fig. \protect\ref{fig:c_gamma_0.1}, but for $\protect%
\gamma =0.9$.}
\label{fig:c_gamma_0.9}
\end{figure}

At $k\neq 0$, the numerically found solutions are still close to the the
strongly asymmetric analytical expressions given by Eqs. (\ref{0}) and (\ref%
{Vexact}) for $k=0$, provided that $|k|$ is small enough, see Fig. \ref%
{fig:c_gamma_0.2_k_0.2} for $\gamma =0.2$ and $k=0.2$. However, at larger $k$%
, such as $k=0.8$ with the same $\gamma =0.2$ [note that $k=0.8$ falls into
the bandgap (\ref{gap}) in this case], the numerical solution for the $V$
component strongly deviates from the analytical expression given by Eq. (\ref%
{V}) for $k=0$, while the $U$ component is still close to the simple
analytical form (\ref{0}), see Fig. \ref{fig:c_gamma_0.2_k_0.8}.
\begin{figure}[tbp]
\centering
\includegraphics[width=0.5\textwidth]{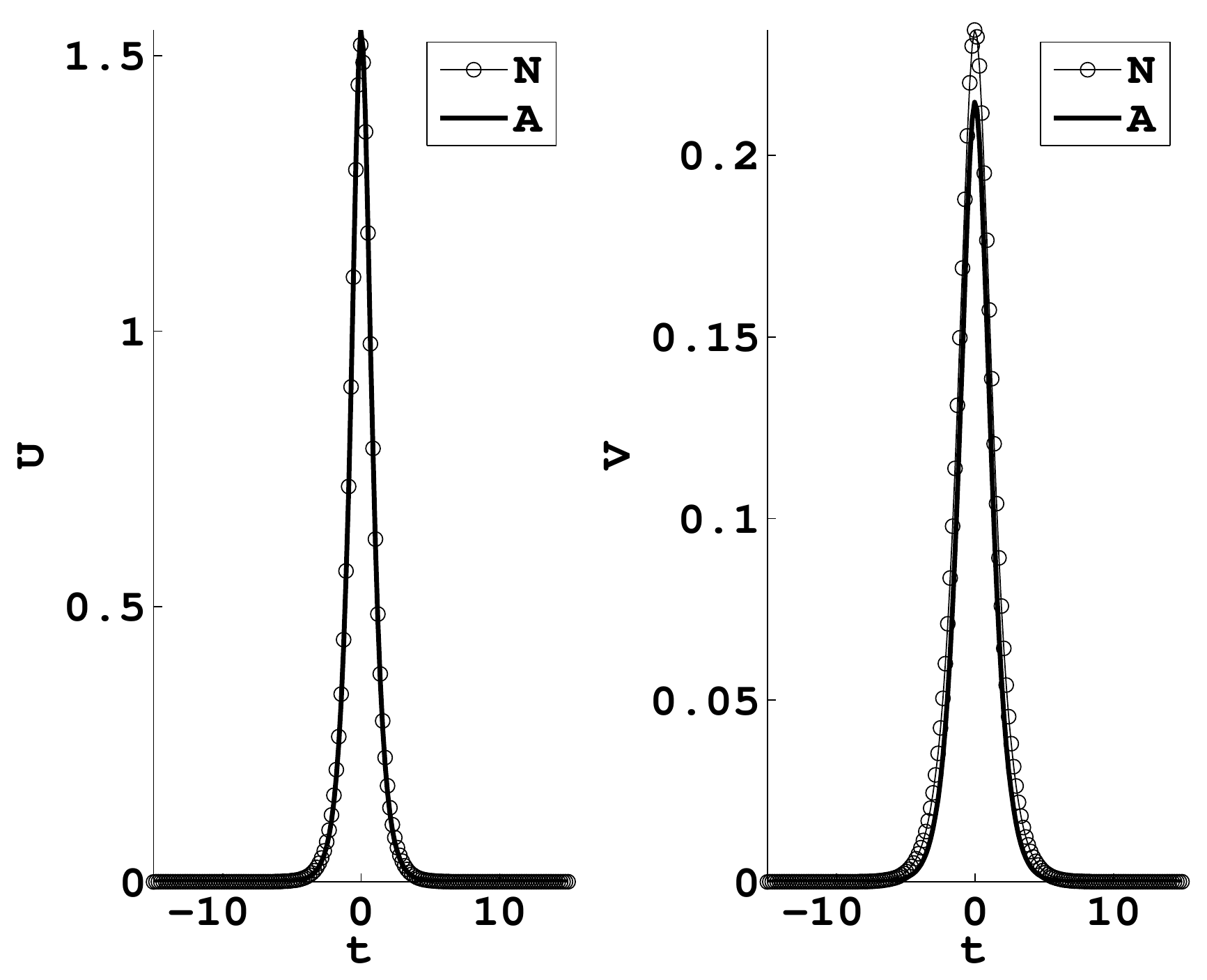}
\caption{The same as in Fig. \protect\ref{fig:c_gamma_0.1}, but for $\protect%
\gamma =0.2$, and the numerical solution taken for $k=0.2$ [recall that
analytical expression (\protect\ref{Vexact}) pertains to $k=0$]. }
\label{fig:c_gamma_0.2_k_0.2}
\end{figure}
\begin{figure}[tbp]
\centering
\includegraphics[width=0.5\textwidth]{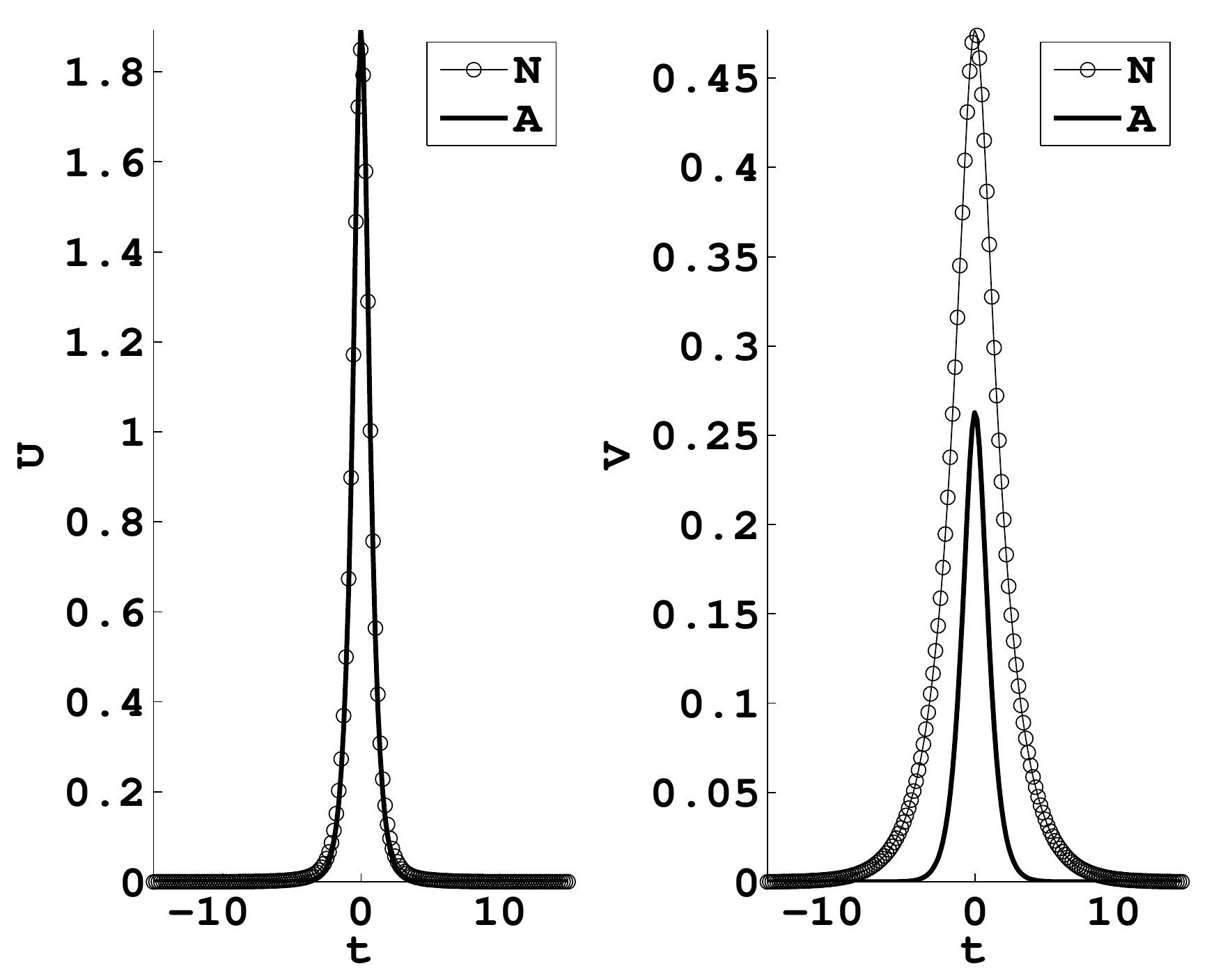}
\caption{The same as in Fig. \protect\ref{fig:c_gamma_0.2_k_0.2}, but for
the numerical solution taken for $k=0.8$.}
\label{fig:c_gamma_0.2_k_0.8}
\end{figure}

An analytical solution for strongly asymmetric gap solitons (corresponding
to $\gamma \ll 1$) was also obtained above in the form of Eqs. (\ref{0}) and
(\ref{3/5}) for $k=-(3/5)q$. For $\gamma =0.1$, this solution is virtually
identical to its numerically found counterpart, as shown in Fig. \ref%
{fig:c_gamma_0.1_k_3_5}.

Finally, the stability of the numerically generated gap solitons was tested
by using them as initial conditions in direct simulations of Eqs. (\ref{u})
and (\ref{v}). The results, not shown here in detail, demonstrate that all
the tested examples of gap solitons are stable, both at the center of the
gap, $k=0$, and off the center, including values of the coupling constant
(such as $\gamma =0.9$) which are close to the symmetry-breaking threshold (%
\ref{thr}).

\section{Conclusions}

The objective of this work is to introduce a model of a dual-core waveguide
which may implement an optical system featuring the $\mathcal{CP}$ symmetry.
Essential ingredients of the model are opposite GVD signs in the two cores,
a phase-velocity mismatch between them, and the linear coupling of the
gain-loss type, which is possible if the waveguide is embedded into an
active medium, or may be provided by the propagation in the $\chi ^{(2)}$
medium of the Type-II (three-wave) type, neglecting the depletion of the SH
(second-harmonic) pump. Nonlinear cubic terms, which destroy the symmetry,
were considered as well (in the case of the $\chi ^{(2)}$ medium, they are
different from those considered here). It is predicted in an approximate
analytical form and demonstrated numerically that the linear system gives
rise to expanding Gaussian pulses. Relatively weak nonlinearity produces an
essential effect, building broad oscillatory quasi-solitons, which are
destroyed in direct simulations if the nonlinearity is too strong. Further,
the analysis predicts a general family of stationary gap solitons in the
nonlinear system, that have been also found and checked for the stability in
the numerical form, the broad solitary pulses being a limit case of the gap
solitons near the bottom edge of the bandgap.

The analysis can be continued by considering higher-order modes [it is well
known that linear Schr\"{o}dinger equations (\ref{slow}) give rise to higher
modes in the form of Hermite-Gauss wave functions] and interactions between
fundamental solitons in the nonlinear version of the system. On the other
hand, it was mentioned above that Eq. (\ref{slow}) for $\tilde{v}$ suggests
the existence of dark solitons in the present system, which is an
interesting issue too. Still another possibility is to consider the
different form of the cubic nonlinearity, corresponding to the underlying $%
\chi ^{(2)}$ system. A challenging extension is to consider a
two-dimensional version of the model, which may be based on a dual-core
planar waveguide embedded into the active medium.


\begin{thebibliography}{{Skotiniotis} et~al.(2012){Skotiniotis}, {Durham}, {%
Toloui}, and {Sanders}}
\bibitem{} \expandafter\ifx\csname natexlab\endcsname\relax

\fi
\expandafter\ifx\csname bibnamefont\endcsname\relax

\fi
\expandafter\ifx\csname bibfnamefont\endcsname\relax

\fi
\expandafter\ifx\csname citenamefont\endcsname\relax

\fi
\expandafter\ifx\csname url\endcsname\relax

\fi
\expandafter\ifx\csname urlprefix\endcsname\relax

\fi
\providecommand{\bibinfo}[2]{#2} \providecommand{\eprint}[2][]{\url{#2}}

\bibitem[Bernabeu(2011)]{1742-6596-335-1-012011} \bibinfo{author}{%
\bibfnamefont{J.}~\bibnamefont{Bernabeu}}, in \emph{%
\bibinfo{booktitle}{Journal of Physics: Conference Series}} (%
\bibinfo{organization}{IOP Publishing}, \bibinfo{year}{2011}), vol. %
\bibinfo{volume}{335}, p. \bibinfo{pages}{012011}.

\bibitem[Aguilar-Arevalo et~al.(2013)Aguilar-Arevalo, Anderson, Bazarko,
Brice, Brown, Bugel, Cao, Coney, Conrad, Cox et~al.]{AguilarArevalo20131303} %
\bibinfo{author}{\bibfnamefont{A.}~\bibnamefont{Aguilar-Arevalo}}, %
\bibinfo{author}{\bibfnamefont{C.}~\bibnamefont{Anderson}}, %
\bibinfo{author}{\bibfnamefont{A.}~\bibnamefont{Bazarko}}, %
\bibinfo{author}{\bibfnamefont{S.}~\bibnamefont{Brice}}, \bibinfo{author}{%
\bibfnamefont{B.}~\bibnamefont{Brown}}, \bibinfo{author}{\bibfnamefont{L.}~%
\bibnamefont{Bugel}}, \bibinfo{author}{\bibfnamefont{J.}~\bibnamefont{Cao}}, %
\bibinfo{author}{\bibfnamefont{L.}~\bibnamefont{Coney}}, \bibinfo{author}{%
\bibfnamefont{J.}~\bibnamefont{Conrad}}, \bibinfo{author}{\bibfnamefont{D.}~%
\bibnamefont{Cox}}, et~al., \bibinfo{journal}{Physics Letters B} \textbf{%
\bibinfo{volume}{718}}, \bibinfo{pages}{1303 } (\bibinfo{year}{2013}).

\bibitem[Yaffe(2013)]{particle}
\bibinfo{author}{\bibfnamefont{L.~G.}
\bibnamefont{Yaffe}}, \emph{\bibinfo{title}{Particles and Symmetries}} (%
\bibinfo{publisher}{University of Washington}, \bibinfo{year}{2013}).

\bibitem[Bender(2003)]{Ham}
\bibinfo{author}{\bibfnamefont{C.~M.}
\bibnamefont{Bender}}, \bibinfo{journal}{Annales de l'institut Fourier}
\textbf{\bibinfo{volume}{53}}, \bibinfo{pages}{997} (\bibinfo{year}{2003}), %
\eprint{1210.0208}.

\bibitem[Bender et~al.(2012)Bender, Fring, G\"{u}nther, and Jones]{Bender}
\bibinfo{author}{\bibfnamefont{C.}~%
\bibnamefont{Bender}}, \bibinfo{author}{\bibfnamefont{A.}~%
\bibnamefont{Fring}}, \bibinfo{author}{\bibfnamefont{U.}~%
\bibnamefont{Gunther}}, and \bibinfo{author}{\bibfnamefont{H.}~%
\bibnamefont{Jones}},
\bibinfo{journal}{Journal of Physics A: Mathematical
and Theoretical} \textbf{\bibinfo{volume}{45}}, \bibinfo{pages}{440301} (%
\bibinfo{year}{2012}).

\bibitem[Ruschhaupt et~al.(2005)Ruschhaupt, Delgado, and Muga]{Muga} %
\bibinfo{author}{\bibfnamefont{A.}~\bibnamefont{Ruschhaupt}}, %
\bibinfo{author}{\bibfnamefont{F.}~\bibnamefont{Delgado}}, and %
\bibinfo{author}{\bibfnamefont{J.~G.} \bibnamefont{Muga}}, %
\bibinfo{journal}{Journal of Physics A: Mathematical and General} \textbf{%
\bibinfo{volume}{38}}, \bibinfo{pages}{L171} (\bibinfo{year}{2005}).

\bibitem[Musslimani et~al.(2008)Musslimani, Makris, El-Ganainy, and
Christodoulides]{Muga1}
\bibinfo{author}{\bibfnamefont{Z.~H.}
\bibnamefont{Musslimani}},
\bibinfo{author}{\bibfnamefont{K.~G.}
\bibnamefont{Makris}}, \bibinfo{author}{\bibfnamefont{R.}~%
\bibnamefont{El-Ganainy}}, and
\bibinfo{author}{\bibfnamefont{D.~N.}
  \bibnamefont{Christodoulides}}, \bibinfo{journal}{Phys. Rev. Lett.}
\textbf{\bibinfo{volume}{100}}, \bibinfo{pages}{030402} (\bibinfo{year}{2008}%
).

\bibitem[Makris et~al.(2008)Makris, El-Ganainy, Christodoulides, and
Musslimani]{Muga2}
\bibinfo{author}{\bibfnamefont{K.~G.}
\bibnamefont{Makris}}, \bibinfo{author}{\bibfnamefont{R.}~%
\bibnamefont{El-Ganainy}},
\bibinfo{author}{\bibfnamefont{D.~N.}
\bibnamefont{Christodoulides}}, and
\bibinfo{author}{\bibfnamefont{Z.~H.}
  \bibnamefont{Musslimani}}, \bibinfo{journal}{Phys. Rev. Lett.} \textbf{%
\bibinfo{volume}{100}}, \bibinfo{pages}{103904} (\bibinfo{year}{2008}).

\bibitem[Longhi(2010)]{Muga3} \bibinfo{author}{\bibfnamefont{S.}~%
\bibnamefont{Longhi}}, \bibinfo{journal}{Phys. Rev. A} \textbf{%
\bibinfo{volume}{81}}, \bibinfo{pages}{022102} (\bibinfo{year}{2010}).

\bibitem[Lin et~al.(2011)Lin, Ramezani, Eichelkraut, Kottos, Cao, and
Christodoulides]{Muga4} \bibinfo{author}{\bibfnamefont{Z.}~\bibnamefont{Lin}}%
, \bibinfo{author}{\bibfnamefont{H.}~\bibnamefont{Ramezani}}, %
\bibinfo{author}{\bibfnamefont{T.}~\bibnamefont{Eichelkraut}}, %
\bibinfo{author}{\bibfnamefont{T.}~\bibnamefont{Kottos}}, %
\bibinfo{author}{\bibfnamefont{H.}~\bibnamefont{Cao}}, and %
\bibinfo{author}{\bibfnamefont{D.~N.} \bibnamefont{Christodoulides}}, %
\bibinfo{journal}{Phys. Rev. Lett.} \textbf{\bibinfo{volume}{106}}, %
\bibinfo{pages}{213901} (\bibinfo{year}{2011}).

\bibitem[Zhu et~al.(2011)Zhu, Wang, Zheng, Li, and He]{Muga5} %
\bibinfo{author}{\bibfnamefont{X.}~\bibnamefont{Zhu}}, \bibinfo{author}{%
\bibfnamefont{H.}~\bibnamefont{Wang}},
\bibinfo{author}{\bibfnamefont{L.-X.}
\bibnamefont{Zheng}}, \bibinfo{author}{\bibfnamefont{H.}~\bibnamefont{Li}},
and \bibinfo{author}{\bibfnamefont{Y.-J.} \bibnamefont{He}}, %
\bibinfo{journal}{Opt. Lett.} \textbf{\bibinfo{volume}{36}}, %
\bibinfo{pages}{2680} (\bibinfo{year}{2011}).

\bibitem[Makris et~al.(2011)Makris, El-Ganainy, Christodoulides, and
Musslimani]{review} \bibinfo{author}{\bibfnamefont{K.}~\bibnamefont{Makris}}%
, \bibinfo{author}{\bibfnamefont{R.}~\bibnamefont{El-Ganainy}}, %
\bibinfo{author}{\bibfnamefont{D.}~\bibnamefont{Christodoulides}}, and %
\bibinfo{author}{\bibfnamefont{Z.}~\bibnamefont{Musslimani}}, %
\bibinfo{journal}{International Journal of Theoretical Physics} \textbf{%
\bibinfo{volume}{50}}, \bibinfo{pages}{1019} (\bibinfo{year}{2011}).

\bibitem[Ruter et~al.(2010{a})Ruter, Makris, El-Ganainy, Christodoulides,
Segev, and Kip]{Kip}
\bibinfo{author}{\bibfnamefont{C.~E.}
\bibnamefont{Ruter}},
\bibinfo{author}{\bibfnamefont{K.~G.}
\bibnamefont{Makris}}, \bibinfo{author}{\bibfnamefont{R.}~%
\bibnamefont{El-Ganainy}},
\bibinfo{author}{\bibfnamefont{D.~N.}
\bibnamefont{Christodoulides}}, \bibinfo{author}{\bibfnamefont{M.}~%
\bibnamefont{Segev}}, and \bibinfo{author}{\bibfnamefont{D.}~%
\bibnamefont{Kip}}, \textbf{\bibinfo{volume}{6}}, \bibinfo{pages}{192} (%
\bibinfo{year}{2010}{\natexlab{a}}).

\bibitem[Feng et~al.(2011)Feng, Ayache, Huang, Xu, Lu, Chen, Fainman, and
Scherer]{Feng05082011} \bibinfo{author}{\bibfnamefont{L.}~\bibnamefont{Feng}}%
, \bibinfo{author}{\bibfnamefont{M.}~\bibnamefont{Ayache}}, %
\bibinfo{author}{\bibfnamefont{J.}~\bibnamefont{Huang}}, \bibinfo{author}{%
\bibfnamefont{Y.-L.} \bibnamefont{Xu}}, \bibinfo{author}{%
\bibfnamefont{M.-H.} \bibnamefont{Lu}}, \bibinfo{author}{%
\bibfnamefont{Y.-F.} \bibnamefont{Chen}}, \bibinfo{author}{%
\bibfnamefont{Y.}~\bibnamefont{Fainman}}, and \bibinfo{author}{%
\bibfnamefont{A.}~\bibnamefont{Scherer}}, \bibinfo{journal}{Science} \textbf{%
\bibinfo{volume}{333}}, \bibinfo{pages}{729} (\bibinfo{year}{2011}).

\bibitem[Regensburger et~al.(2012)Regensburger, Bersch, Miri, Onishchukov,
Christodoulides, and Peschel]{citeulike:11031904} \bibinfo{author}{%
\bibfnamefont{A.}~\bibnamefont{Regensburger}}, \bibinfo{author}{%
\bibfnamefont{C.}~\bibnamefont{Bersch}}, \bibinfo{author}{%
\bibfnamefont{M.-A.} \bibnamefont{Miri}}, \bibinfo{author}{%
\bibfnamefont{G.}~\bibnamefont{Onishchukov}}, \bibinfo{author}{%
\bibfnamefont{D.~N.} \bibnamefont{Christodoulides}}, and \bibinfo{author}{%
\bibfnamefont{U.}~\bibnamefont{Peschel}}, \bibinfo{journal}{Nature} \textbf{%
\bibinfo{volume}{488}}, \bibinfo{pages}{167} (\bibinfo{year}{2012}).

\bibitem[R\"{u}ter et~al.(2009)R\"{u}ter, Kip, Makris, Christodoulides,
Peleg, and Segev]{Ruter:09}
\bibinfo{author}{\bibfnamefont{C.~E.}
\bibnamefont{R\"{u}ter}}, \bibinfo{author}{\bibfnamefont{D.}~%
\bibnamefont{Kip}},
\bibinfo{author}{\bibfnamefont{K.~G.}
\bibnamefont{Makris}},
\bibinfo{author}{\bibfnamefont{D.~N.}
\bibnamefont{Christodoulides}}, \bibinfo{author}{\bibfnamefont{O.}~%
\bibnamefont{Peleg}}, and \bibinfo{author}{\bibfnamefont{M.}~%
\bibnamefont{Segev}}, in \emph{%
\bibinfo{booktitle}{Conference on Lasers and
  Electro-Optics/International Quantum Electronics Conference}} (%
\bibinfo{publisher}{Optical Society of America}, \bibinfo{year}{2009}), p. %
\bibinfo{pages}{ITuF2}.

\bibitem[Yu and Fan(2009)]{10.1117/12.807739} \bibinfo{author}{%
\bibfnamefont{Z.}~\bibnamefont{Yu}} and \bibinfo{author}{\bibfnamefont{S.}~%
\bibnamefont{Fan}} (\bibinfo{year}{2009}), vol. \bibinfo{volume}{7220}, pp. %
\bibinfo{pages}{72200W--72200W--7}.

\bibitem[Zheng et~al.(2010)Zheng, Christodoulides, Fleischmann, and Kottos]%
{PhysRevA.82.010103}
\bibinfo{author}{\bibfnamefont{M.~C.}
\bibnamefont{Zheng}},
\bibinfo{author}{\bibfnamefont{D.~N.}
\bibnamefont{Christodoulides}}, \bibinfo{author}{\bibfnamefont{R.}~%
\bibnamefont{Fleischmann}}, and \bibinfo{author}{\bibfnamefont{T.}~%
\bibnamefont{Kottos}}, \bibinfo{journal}{Phys. Rev. A} \textbf{%
\bibinfo{volume}{82}}, \bibinfo{pages}{010103} (\bibinfo{year}{2010}).

\bibitem[{Razzari} and {Morandotti}(2012)]{2012Natur.488..163R} %
\bibinfo{author}{\bibfnamefont{L.}~\bibnamefont{{Razzari}}} and %
\bibinfo{author}{\bibfnamefont{R.}~\bibnamefont{{Morandotti}}}, %
\bibinfo{journal}{\nat} \textbf{\bibinfo{volume}{488}}, \bibinfo{pages}{163}
(\bibinfo{year}{2012}).

\bibitem[Bender et~al.(2013)Bender, Factor, Bodyfelt, Ramezani,
Christodoulides, Ellis, and Kottos]{Kottos} \bibinfo{author}{%
\bibfnamefont{N.}~\bibnamefont{Bender}}, \bibinfo{author}{\bibfnamefont{S.}~%
\bibnamefont{Factor}},
\bibinfo{author}{\bibfnamefont{J.~D.}
\bibnamefont{Bodyfelt}}, \bibinfo{author}{\bibfnamefont{H.}~%
\bibnamefont{Ramezani}},
\bibinfo{author}{\bibfnamefont{D.~N.}
\bibnamefont{Christodoulides}},
\bibinfo{author}{\bibfnamefont{F.~M.}
\bibnamefont{Ellis}}, and \bibinfo{author}{\bibfnamefont{T.}~%
\bibnamefont{Kottos}}, \bibinfo{journal}{Phys. Rev. Lett.} \textbf{%
\bibinfo{volume}{110}}, \bibinfo{pages}{234101} (\bibinfo{year}{2013}).

\bibitem[Abdullaev et~al.(2011)Abdullaev, {Konotop}, Ogren, and {Sorensen}]%
{PTsolitons} \bibinfo{author}{\bibfnamefont{F.~K.} \bibnamefont{Abdullaev}}, %
\bibinfo{author}{\bibfnamefont{V.~V.} \bibnamefont{{Konotop}}}, %
\bibinfo{author}{\bibfnamefont{M.}~\bibnamefont{Ogren}}, and %
\bibinfo{author}{\bibfnamefont{M.~P.} \bibnamefont{{Sorensen}}}, %
\bibinfo{journal}{Optics Letters} \textbf{\bibinfo{volume}{36}}, %
\bibinfo{pages}{4566} (\bibinfo{year}{2011}), \eprint{1111.1310}.

\bibitem[Li et~al.(2012)Li, Liu, and Dong]{PTsolitons1} \bibinfo{author}{%
\bibfnamefont{C.}~\bibnamefont{Li}}, \bibinfo{author}{\bibfnamefont{H.}~%
\bibnamefont{Liu}}, and \bibinfo{author}{\bibfnamefont{L.}~%
\bibnamefont{Dong}}, \bibinfo{journal}{Opt. Express} \textbf{%
\bibinfo{volume}{20}}, \bibinfo{pages}{16823} (\bibinfo{year}{2012}).

\bibitem[Suchkov et~al.(2011)Suchkov, Malomed, Dmitriev, and Kivshar]%
{PTsolitons2} \bibinfo{author}{\bibfnamefont{S.~V.} \bibnamefont{Suchkov}}, %
\bibinfo{author}{\bibfnamefont{B.~A.} \bibnamefont{Malomed}}, %
\bibinfo{author}{\bibfnamefont{S.~V.} \bibnamefont{Dmitriev}}, and
\bibinfo{author}{\bibfnamefont{Y.~S.}
  \bibnamefont{Kivshar}}, \bibinfo{journal}{Phys. Rev. E} \textbf{%
\bibinfo{volume}{84}}, \bibinfo{pages}{046609} (\bibinfo{year}{2011}).

\bibitem[Nixon et~al.(2012)Nixon, Ge, and Yang]{PTsolitons3} %
\bibinfo{author}{\bibfnamefont{S.}~\bibnamefont{Nixon}}, \bibinfo{author}{%
\bibfnamefont{L.}~\bibnamefont{Ge}}, and \bibinfo{author}{\bibfnamefont{J.}~%
\bibnamefont{Yang}}, \bibinfo{journal}{Phys. Rev. A} \textbf{%
\bibinfo{volume}{85}}, \bibinfo{pages}{023822} (\bibinfo{year}{2012}).

\bibitem[Zezyulin and Konotop(2012)]{PTsolitons4} \bibinfo{author}{%
\bibfnamefont{D.~A.} \bibnamefont{Zezyulin}} and \bibinfo{author}{%
\bibfnamefont{V.~V.} \bibnamefont{Konotop}},
\bibinfo{journal}{Phys. Rev.
Lett.} \textbf{\bibinfo{volume}{108}}, \bibinfo{pages}{213906} (%
\bibinfo{year}{2012}).

\bibitem[Leykam et~al.(2013)Leykam, Konotop, and Desyatnikov]{PTsolitons5} %
\bibinfo{author}{\bibfnamefont{D.}~\bibnamefont{Leykam}}, %
\bibinfo{author}{\bibfnamefont{V.~V.} \bibnamefont{Konotop}}, and
\bibinfo{author}{\bibfnamefont{A.~S.}
  \bibnamefont{Desyatnikov}}, \bibinfo{journal}{Opt. Lett.} \textbf{%
\bibinfo{volume}{38}}, \bibinfo{pages}{371} (\bibinfo{year}{2013}).

\bibitem[Driben and Malomed(2011)]{couplers} \bibinfo{author}{%
\bibfnamefont{R.}~\bibnamefont{Driben}} and \bibinfo{author}{%
\bibfnamefont{B.~A.} \bibnamefont{Malomed}}, \bibinfo{journal}{Opt. Lett.}
\textbf{\bibinfo{volume}{36}}, \bibinfo{pages}{4323} (\bibinfo{year}{2011}).

\bibitem[Alexeeva et~al.(2012)Alexeeva, Barashenkov, Sukhorukov, and Kivshar]%
{couplers1} \bibinfo{author}{\bibfnamefont{N.~V.} \bibnamefont{Alexeeva}}, %
\bibinfo{author}{\bibfnamefont{I.~V.} \bibnamefont{Barashenkov}}, %
\bibinfo{author}{\bibfnamefont{A.~A.} \bibnamefont{Sukhorukov}}, and
\bibinfo{author}{\bibfnamefont{Y.~S.}
  \bibnamefont{Kivshar}}, \bibinfo{journal}{Phys. Rev. A} \textbf{%
\bibinfo{volume}{85}}, \bibinfo{pages}{063837} (\bibinfo{year}{2012}).

\bibitem[{Driben} and {Malomed}(2011)]{coupler-management} %
\bibinfo{author}{\bibfnamefont{R.}~\bibnamefont{{Driben}}} and %
\bibinfo{author}{\bibfnamefont{B.~A.} \bibnamefont{{Malomed}}}, %
\bibinfo{journal}{EPL (Europhysics Letters)} \textbf{\bibinfo{volume}{96}}, %
\bibinfo{pages}{51001} (\bibinfo{year}{2011}), \eprint{1110.2409}.

\bibitem[Moreira et~al.(2012)Moreira, Abdullaev, Konotop, and Yulin]{VVK1} %
\bibinfo{author}{\bibfnamefont{F.~C.} \bibnamefont{Moreira}}, %
\bibinfo{author}{\bibfnamefont{F.~K.} \bibnamefont{Abdullaev}}, %
\bibinfo{author}{\bibfnamefont{V.~V.} \bibnamefont{Konotop}}, and %
\bibinfo{author}{\bibfnamefont{A.~V.} \bibnamefont{Yulin}}, %
\bibinfo{journal}{Phys. Rev. A} \textbf{\bibinfo{volume}{86}}, %
\bibinfo{pages}{053815} (\bibinfo{year}{2012}).

\bibitem[Moreira et~al.(2013)Moreira, Konotop, and Malomed]{VVK2} %
\bibinfo{author}{\bibfnamefont{F.~C.} \bibnamefont{Moreira}}, %
\bibinfo{author}{\bibfnamefont{V.~V.} \bibnamefont{Konotop}}, and
\bibinfo{author}{\bibfnamefont{B.~A.}
  \bibnamefont{Malomed}}, \bibinfo{journal}{Phys. Rev. A} \textbf{%
\bibinfo{volume}{87}}, \bibinfo{pages}{013832} (\bibinfo{year}{2013}).

\bibitem[Li et~al.(2013)Li, Zezyulin, Kevrekidis, Konotop, and Abdullaev]%
{VVK3} \bibinfo{author}{\bibfnamefont{K.}~\bibnamefont{Li}}, %
\bibinfo{author}{\bibfnamefont{D.~A.} \bibnamefont{Zezyulin}}, %
\bibinfo{author}{\bibfnamefont{P.~G.} \bibnamefont{Kevrekidis}}, %
\bibinfo{author}{\bibfnamefont{V.~V.} \bibnamefont{Konotop}}, and
\bibinfo{author}{\bibfnamefont{F.~K.}
  \bibnamefont{Abdullaev}}, \bibinfo{journal}{Phys. Rev. A} \textbf{%
\bibinfo{volume}{88}}, \bibinfo{pages}{053820} (\bibinfo{year}{2013}).

\bibitem[{Skotiniotis} et~al.(2012){Skotiniotis}, {Durham}, {Toloui}, and {%
Sanders}]{2012arXiv1201.1594S} \bibinfo{author}{\bibfnamefont{M.}~%
\bibnamefont{{Skotiniotis}}},
\bibinfo{author}{\bibfnamefont{I.~T.}
\bibnamefont{{Durham}}}, \bibinfo{author}{\bibfnamefont{B.}~%
\bibnamefont{{Toloui}}}, and
\bibinfo{author}{\bibfnamefont{B.~C.}
\bibnamefont{{Sanders}}}, \bibinfo{journal}{ArXiv e-prints} (%
\bibinfo{year}{2012}), \eprint{1201.1594}.

\bibitem[Srednicki(2007)]{srednicki2007quantum} \bibinfo{author}{%
\bibfnamefont{M.}~\bibnamefont{Srednicki}}, \emph{%
\bibinfo{title}{Quantum
Field Theory}} (\bibinfo{publisher}{Cambridge
  University Press}, \bibinfo{year}{2007}).

\bibitem[Kursunogammalu et~al.(2013)Kursunogammalu, Mintz, and Perlmutter]%
{kursunogammalu2013confluence} \bibinfo{author}{\bibfnamefont{B.}~%
\bibnamefont{Kursunogammalu}}, \bibinfo{author}{\bibfnamefont{S.}~%
\bibnamefont{Mintz}}, and \bibinfo{author}{\bibfnamefont{A.}~%
\bibnamefont{Perlmutter}}, \emph{%
\bibinfo{title}{Confluence of Cosmology, Massive Neutrinos, Elementary
  Particles, and Gravitation}} (\bibinfo{publisher}{Springer US}, %
\bibinfo{year}{2013}).

\bibitem[{Chaichian} et~al.(2013){Chaichian}, {Fujikawa}, and {Tureanu}]%
{2013PhLB..718.1500C} \bibinfo{author}{\bibfnamefont{M.}~%
\bibnamefont{{Chaichian}}}, \bibinfo{author}{\bibfnamefont{K.}~%
\bibnamefont{{Fujikawa}}}, and \bibinfo{author}{\bibfnamefont{A.}~%
\bibnamefont{{Tureanu}}}, \bibinfo{journal}{Physics Letters B} \textbf{%
\bibinfo{volume}{718}}, \bibinfo{pages}{1500} (\bibinfo{year}{2013}), %
\eprint{1210.0208}.

\bibitem[Kartashov et~al.(2014)Kartashov, Konotop, and Zezyulin]{something} %
\bibinfo{author}{\bibfnamefont{Y.~V.} \bibnamefont{Kartashov}}, %
\bibinfo{author}{\bibfnamefont{V.~V.} \bibnamefont{Konotop}}, and
\bibinfo{author}{\bibfnamefont{D.~A.}
  \bibnamefont{Zezyulin}}, \bibinfo{journal}{EPL (Europhysics Letters)}
\textbf{\bibinfo{volume}{107}}, \bibinfo{pages}{50002} (\bibinfo{year}{2014}%
).

\bibitem[Boardman and Xie(1994)]{Kaup}
\bibinfo{author}{\bibfnamefont{A.~D.}
\bibnamefont{Boardman}} and \bibinfo{author}{\bibfnamefont{K.}~%
\bibnamefont{Xie}}, \bibinfo{journal}{Phys. Rev. A} \textbf{%
\bibinfo{volume}{50}}, \bibinfo{pages}{1851} (\bibinfo{year}{1994}).

\bibitem[Kaup and Malomed(1998)]{Kaup1} \bibinfo{author}{%
\bibfnamefont{D.~J.} \bibnamefont{Kaup}} and \bibinfo{author}{%
\bibfnamefont{B.~A.} \bibnamefont{Malomed}},
\bibinfo{journal}{J. Opt. Soc.
Am. B} \textbf{\bibinfo{volume}{15}}, \bibinfo{pages}{2838} (%
\bibinfo{year}{1998}).

\bibitem[Dana et~al.(2014)Dana, Malomed, and Bahabad]{Dana:14} %
\bibinfo{author}{\bibfnamefont{B.}~\bibnamefont{Dana}}, \bibinfo{author}{%
\bibfnamefont{B.~A.} \bibnamefont{Malomed}}, and \bibinfo{author}{%
\bibfnamefont{A.}~\bibnamefont{Bahabad}}, \bibinfo{journal}{Opt. Lett.}
\textbf{\bibinfo{volume}{39}}, \bibinfo{pages}{2175} (\bibinfo{year}{2014}).

\bibitem{Maim1} N. M. Litchinitser, I. R. Gabitov, and A. I. Maimistov,
Phys. Rev. Lett. \textbf{99}, 113902 (2007).

\bibitem{Maim2} A. I. Maimistov and E.V. Kazantseva, Optika i Spektroskopiya
\textbf{112}, 291 (2012) [English translation: Optics and Spectroscopy
\textbf{112}, 264 (2012)].

\bibitem{Maim3} A. A. Dovgiy and A. I. Maimistov, Optika i Spektroskopiya
\textbf{116}, 673 (2014) [English translation: Optics and Spectroscopy
\textbf{116}, 626 (2014)].

\bibitem[Alexeeva et~al.(2014)Alexeeva, Barashenkov, Rayanov, and Flach]%
{Barash} \bibinfo{author}{\bibfnamefont{N.~V.} \bibnamefont{Alexeeva}}, %
\bibinfo{author}{\bibfnamefont{I.~V.} \bibnamefont{Barashenkov}}, %
\bibinfo{author}{\bibfnamefont{K.}~\bibnamefont{Rayanov}}, and %
\bibinfo{author}{\bibfnamefont{S.}~\bibnamefont{Flach}}, %
\bibinfo{journal}{Phys. Rev. A} \textbf{\bibinfo{volume}{89}}, %
\bibinfo{pages}{013848} (\bibinfo{year}{2014}).

\bibitem{VVK} D. A. Zezyulin, V. V. Konotop, and F. Kh. Abdullaev, Opt.
Lett. \textbf{37}, 3930 (2012).

\bibitem[Agrawal(2001)]{agrawal2001nonlinear} \bibinfo{author}{%
\bibfnamefont{G.}~\bibnamefont{Agrawal}}, \emph{%
\bibinfo{title}{Nonlinear
Fiber Optics}}, Optics and Photonics (\bibinfo{publisher}{Elsevier Science}, %
\bibinfo{year}{2001}).

\bibitem[Tsoy et~al.(2014)Tsoy, Allayarov, and Abdullaev]{non-PT1} %
\bibinfo{author}{\bibfnamefont{E.~N.} \bibnamefont{Tsoy}}, %
\bibinfo{author}{\bibfnamefont{I.~M.} \bibnamefont{Allayarov}}, and
\bibinfo{author}{\bibfnamefont{F.~K.}
  \bibnamefont{Abdullaev}}, \bibinfo{journal}{Opt. Lett.} \textbf{%
\bibinfo{volume}{39}}, \bibinfo{pages}{4215} (\bibinfo{year}{2014}).

\bibitem[Konotop and Zezyulin(2014)]{non-PT2} \bibinfo{author}{%
\bibfnamefont{V.~V.} \bibnamefont{Konotop}} and \bibinfo{author}{%
\bibfnamefont{D.~A.} \bibnamefont{Zezyulin}}, \bibinfo{journal}{Opt. Lett.}
\textbf{\bibinfo{volume}{39}}, \bibinfo{pages}{5535} (\bibinfo{year}{2014}).

\bibitem[{Nixon} and {Yang}(2014)]{non-PT3} \bibinfo{author}{%
\bibfnamefont{S.}~\bibnamefont{{Nixon}}} and \bibinfo{author}{%
\bibfnamefont{J.}~\bibnamefont{{Yang}}}, \bibinfo{journal}{ArXiv e-prints} (%
\bibinfo{year}{2014}), \eprint{1412.6113}.

\bibitem[Ruter et~al.(2010{b})Ruter, Makris, El-Ganainy, Christodoulides,
Segev, and Kip]{nature}
\bibinfo{author}{\bibfnamefont{C.~E.}
\bibnamefont{Ruter}},
\bibinfo{author}{\bibfnamefont{K.~G.}
\bibnamefont{Makris}}, \bibinfo{author}{\bibfnamefont{R.}~%
\bibnamefont{El-Ganainy}},
\bibinfo{author}{\bibfnamefont{D.~N.}
\bibnamefont{Christodoulides}}, \bibinfo{author}{\bibfnamefont{M.}~%
\bibnamefont{Segev}}, and \bibinfo{author}{\bibfnamefont{D.}~%
\bibnamefont{Kip}}, \textbf{\bibinfo{volume}{6}}, \bibinfo{pages}{192} (%
\bibinfo{year}{2010}{\natexlab{b}}).

\bibitem{chi2-0} G. I. Stegeman, D. J. Hagan, and L. Torner, Opt. Quant.
Elect. \textbf{28}, 1691 (1996),

\bibitem{chi2-1} and U. Peschel, Progr. Opt. \textbf{41}, 483 (2000).

\bibitem{chi2-2} A. V. Buryak, P. Di Trapani, D. V. Skryabin, and S. Trillo,
Phys. Rep. \textbf{370}, 63 (2002).

\bibitem{chi2-3} H. Suchowski, G. Porat, and A. Arie, Laser Opt. Rev.
\textbf{8}, 333 (1014).

\bibitem[{Kartashov} et~al.(2014){Kartashov}, {Malomed}, and {Torner}]%
{unbreakable}
\bibinfo{author}{\bibfnamefont{Y.~V.}
\bibnamefont{{Kartashov}}},
\bibinfo{author}{\bibfnamefont{B.~A.}
\bibnamefont{{Malomed}}}, and \bibinfo{author}{\bibfnamefont{L.}~%
\bibnamefont{{Torner}}}, \bibinfo{journal}{ArXiv e-prints} (%
\bibinfo{year}{2014}), \eprint{1408.6174}.

\bibitem[de~Sterke and Sipe(1994)]{Sterke} \bibinfo{author}{%
\bibfnamefont{C.~M.} \bibnamefont{de~Sterke}} and \bibinfo{author}{%
\bibfnamefont{J.~E.} \bibnamefont{Sipe}},
\bibinfo{journal}{Progress in
Optics XXXIII} \textbf{\bibinfo{volume}{33}}, \bibinfo{pages}{203} (%
\bibinfo{year}{1994}).

\bibitem[Eggleton et~al.(1999)Eggleton, de~Sterke, and Slusher]{broad-Bragg} %
\bibinfo{author}{\bibfnamefont{B.~J.} \bibnamefont{Eggleton}}, %
\bibinfo{author}{\bibfnamefont{C.~M.} \bibnamefont{de~Sterke}}, and
\bibinfo{author}{\bibfnamefont{R.~E.}
  \bibnamefont{Slusher}}, \bibinfo{journal}{J. Opt. Soc. Am. B} \textbf{%
\bibinfo{volume}{16}}, \bibinfo{pages}{587} (\bibinfo{year}{1999}).

\bibitem[Blit and Malomed(2012)]{Roy} \bibinfo{author}{\bibfnamefont{R.}~%
\bibnamefont{Blit}} and
\bibinfo{author}{\bibfnamefont{B.~A.}
\bibnamefont{Malomed}}, \bibinfo{journal}{Phys. Rev. A} \textbf{%
\bibinfo{volume}{86}}, \bibinfo{pages}{043841} (\bibinfo{year}{2012}).

\bibitem[Anderson(1983)]{Anderson1} \bibinfo{author}{\bibfnamefont{D.}~%
\bibnamefont{Anderson}}, \bibinfo{journal}{Phys. Rev. A} \textbf{%
\bibinfo{volume}{27}}, \bibinfo{pages}{3135} (\bibinfo{year}{1983}).

\bibitem[Malomed(2002)]{Anderson2}
\bibinfo{author}{\bibfnamefont{B.~A.}
\bibnamefont{Malomed}} (\bibinfo{publisher}{Elsevier}, \bibinfo{year}{2002}%
), vol.~\bibinfo{volume}{43} of \emph{\bibinfo{series}{Progress in Optics}},
pp. \bibinfo{pages}{71 -- 193}.

\bibitem[Davis(1984)]{davis1984numerical} \bibinfo{author}{%
\bibfnamefont{M.}~\bibnamefont{Davis}}, \emph{%
\bibinfo{title}{Numerical methods and modeling for chemical
  engineers}}, Wiley series in chemical engineering (%
\bibinfo{publisher}{Wiley}, \bibinfo{year}{1984}).
\end{thebibliography}
\end{document}